\documentclass[aps,prd,reprint,onecolumn,balancelastpage,nofootinbib,preprintnumbers,amsmath,amssymb,superscriptaddress,longbibliography]{revtex4-1}

\usepackage[symbol]{footmisc}
 \usepackage{amsmath,amssymb}
\usepackage{revsymb}
\usepackage{graphicx}
\usepackage{color}  
\newcommand{\nc}{\newcommand*} 
\newcommand{\mU}{{\mathcal{U}}}

\nc{\Eq}[1]{Eq.~\eqref{#1}}     % equation
\nc{\Fig}[1]{Fig.~\ref{#1}}     % figure
\nc{\Table}[1]{Table~\ref{#1}}  % table
\nc{\Sec}[1]{Sec.~\ref{#1}}     % section
\def\({\left(}
\def\){\right)}
\def\[{\left[}
\def\]{\right]}
\def\e{\begin{equation}}
\def\q{\end{equation}}
\def\m{\begin{eqnarray}}
\def\n{\end{eqnarray}}
\begin{document}

\title{Detecting a Gravitational-Wave Background from Inflation with Null Energy Condition Violation: Prospects for Taiji}

%%%%%%%%%%%%%%%%%%%%%%%%%%%%%%%%%%%%%%%%%%%%%%%%%%%%%%%%%%%%%%%%%%%%%%%%%%%%%%%%%%%%%%%%%%%%%%%%
%%%%%%%%%%%%%%%%%%%%%%%%%%%%%%%%%%%% author %%%%%%%%%%%%%%%%%%%%%%%%%%%%%%%%%%%%
\author{Zu-Cheng Chen}
\email{ zuchengchen@hunnu.edu.cn}
\affiliation{Department of Physics and Synergetic Innovation Center for Quantum Effects and Applications, Hunan Normal University, Changsha, Hunan 410081, China}
\affiliation{Institute of Interdisciplinary Studies, Hunan Normal University, Changsha, Hunan 410081, China}
%%%%%%%%%%%%%%%%%%%%%%%%%%%%%%%%%%%%%%%%%%%%%%%%%%%%%%%%%%%%%%%%%
% \author[0000-0002-0297-9633]{Lang~Liu}
\author{Lang~Liu}
\email{Corresponding author: liulang@bnu.edu.cn}
\affiliation{Faculty of Arts and Sciences, Beijing Normal University, Zhuhai 519087, China}
\affiliation{Department of Astronomy, Beijing Normal University, Beijing 100875, China}
\affiliation{Advanced Institute of Natural Sciences, Beijing Normal University, Zhuhai 519087, China}

%%%%%%%%%%%%%%%%%%%%%%%%%%%%%%%%%%%%%%%%%%%%%%%%
\begin{abstract}
The null energy condition (NEC) is a fundamental principle in general relativity, and its violation could leave discernible signatures in gravitational waves (GWs). A violation of the NEC during the primordial era would imprint a blue-tilted spectrum on the stochastic gravitational wave background (SGWB) at nanohertz frequencies, potentially accounting for the recently detected signal by pulsar timing arrays. Remarkably, models of NEC violation during inflation also predict a nearly scale-invariant GW spectrum in the millihertz frequency range, which could be detectable by upcoming space-based GW detectors such as Taiji. The observation of this distinctive spectrum would provide compelling evidence for new physics beyond the standard cosmological paradigm. In this study, we explore Taiji's ability to detect an SGWB arising from NEC violation during inflation, considering various foregrounds and noise sources, including an extragalactic foreground from binary black hole mergers throughout the Universe, a galactic foreground from white dwarf binaries, and the intrinsic noise of the Taiji detector. Employing comprehensive Bayesian parameter estimation techniques to analyze simulated Taiji data, we demonstrate a remarkable precision improvement of three orders of magnitude compared to the NANOGrav 15-year data set for measuring the tensor power spectrum amplitude, $P_{T,2}$, during the second inflationary stage. This substantial enhancement in measurement capabilities underscores Taiji's potential as a powerful probe for investigating the NEC violation in the early Universe.
\end{abstract}
	
\maketitle

%%%%%%%%%%%%%%%%%%%%%%%%%%%%%%%%%%%%%%%%%%%%%%%%%%%%%%%%%%%%%%%%%%%%%%%%%%%%%%%%%%%%%%%%%%%%%%%%
\section{Introduction}

The direct detection of gravitational waves (GWs)~\cite{LIGOScientific:2016aoc} has not only confirmed a major prediction of Einstein's theory of general relativity but also opened a novel observational window into the Universe~\cite{LIGOScientific:2019fpa,LIGOScientific:2020tif,LIGOScientific:2021sio}. While observations have primarily focused on transient GW events from the binary black hole and neutron star mergers, detected by ground-based interferometers like LIGO~\cite{LIGOScientific:2014pky} and Virgo~\cite{VIRGO:2014yos}, these detections have revolutionized our understanding of compact objects and their coalescence dynamics~\cite{LIGOScientific:2018jsj,LIGOScientific:2020kqk,KAGRA:2021duu}. However, an equally promising and more elusive target for GW detectors is the stochastic gravitational-wave background (SGWB), which carries a wealth of information about the early Universe and the fundamental laws of physics.

The landscape of GW astronomy is rapidly advancing, with pulsar timing arrays (PTAs) emerging as a critical tool for probing the low-frequency GW spectrum. Recent announcements from leading PTA collaborations, including NANOGrav~\cite{NANOGrav:2023hde,NANOGrav:2023gor}, PPTA~\cite{Zic:2023gta,Reardon:2023gzh}, EPTA+InPTA~\cite{Antoniadis:2023lym,Antoniadis:2023ott}, and CPTA~\cite{Xu:2023wog}, have provided compelling evidence of a stochastic signal that aligns with the Hellings-Downs spatial correlations~\cite{Hellings:1983fr} expected from an SGWB as predicted by general relativity. These findings mark a significant milestone in the detection of GWs through the timing of highly stable millisecond pulsars.

Despite these achievements, the exact origin of the observed PTA signals remains elusive~\cite{NANOGrav:2023hvm,Antoniadis:2023xlr}, with hypotheses spanning both astrophysical and cosmological sources. The diverse potential origins include supermassive black hole binaries~\cite{NANOGrav:2023hfp,Ellis:2023dgf,Shen:2023pan,Bi:2023tib,Barausse:2023yrx}, as well as more exotic phenomena such as scalar-induced GWs~\cite{Liu:2023ymk,Franciolini:2023pbf,Wang:2023ost,Jin:2023wri,Liu:2023pau,Zhao:2023joc,Yi:2023npi,Harigaya:2023pmw,Balaji:2023ehk,Yi:2023tdk,You:2023rmn,Liu:2023hpw,Domenech:2024rks,Chen:2024twp,Chen:2024fir} that could be linked to primordial black hole formation~\cite{Bhaumik:2023wmw,Liu:2020cds,Bousder:2023ida,Gouttenoire:2023nzr,Huang:2023chx,Depta:2023qst,Wang:2024vfv}, cosmic phase transitions~\cite{Addazi:2023jvg,Athron:2023mer,Zu:2023olm,Jiang:2023qbm,Xiao:2023dbb,Abe:2023yrw,Gouttenoire:2023bqy,An:2023jxf,Chen:2023bms}, domain walls~\cite{Kitajima:2023cek,Blasi:2023sej,Babichev:2023pbf}, and cosmic strings~\cite{Kitajima:2023vre,Ellis:2023tsl,Wang:2023len,Ahmed:2023pjl,Antusch:2023zjk}. Another exciting possibility is that the detected signal is the SGWB from the violation of the null energy condition (NEC) in the early Universe~\cite{Ye:2023tpz}.

The NEC is a fundamental principle in general relativity, asserting that the contraction of any null vector with the energy-momentum tensor yields a non-negative value. However, NEC violation has been hypothesized to explain various early Universe phenomena, such as inflation and primordial black hole formation~\cite{Cai:2017dyi,Kolevatov:2017voe,Ilyas:2020zcb,Ilyas:2020qja,Zhu:2021ggm,Lesnefsky:2022fen,Cai:2023uhc,Chen:2024mwg,Easson:2024fzn}; see the review by~\cite{Rubakov:2014jja}. Recent developments in modified gravity theories, particularly those beyond Horndeski models~\cite{Cai:2012va,Libanov:2016kfc,Kobayashi:2016xpl,Dobre:2017pnt,Cai:2022ori}, have demonstrated the theoretical possibility of stable NEC violations, overcoming pathological instabilities. The violation of the NEC has a significant impact on the generation of primordial GWs in the early Universe, leading to an amplification of the primordial tensor power spectrum~\cite{Cai:2020qpu,Cai:2022nqv}. This amplification is characterized by a blue tilt in the tensor power spectrum, signifying enhanced production of high-frequency GWs. Consequently, the SGWB is expected to exhibit a distinctive pattern that reflects the NEC-violating phase. The detection of this characteristic signature in the SGWB would provide strong evidence for NEC-violating physics and shed light on the nature of gravity at high energies.

Investigating NEC violation through its effect on the SGWB presents a novel approach to probing early Universe physics, complementing other cosmological observations. By comparing the theoretical predictions of NEC-violating models with observational SGWB constraints, valuable insights can be gained into the fundamental laws governing the Universe's earliest stages, enabling the exploration of new physics beyond the standard cosmological paradigm. Remarkably, models of NEC violation during inflation also predict a nearly scale-invariant GW spectrum in the millihertz frequency range, which could be detectable by upcoming space-based GW detectors such as Taiji.
The Taiji program~\cite{Ruan:2018tsw}, China's space-based GW observatory, is poised to contribute significantly to this frontier of physics. With its planned launch in the near future, Taiji will surpass the sensitivity of current ground-based detectors at low frequencies and thus holds the potential to detect the subtle signals of an SGWB. Prospects for Taiji are bolstered by its ability to operate in a frequency band that complements LIGO-Virgo-KAGRA and PTAs, providing a broad observational baseline to search for GW signals.

In this paper, we investigate the SGWB produced by NEC violation during inflation and constrain the model parameters using the NANOGrav 15-year data set. We then analyze simulated Taiji data and apply parameter estimation techniques to quantify the uncertainties in determining the model parameters. This analysis provides insights into the precision and accuracy of parameter estimation for the SGWB produced by NEC violation during inflation, as well as the detection capabilities of Taiji.
The paper is structured as follows. In \Sec{NEC}, we offer a brief review of the SGWB produced by the NEC violation during inflation and the constraints on the model parameters obtained from the NANOGrav 15-year data set. \Sec{model} introduces the various components of the model used in the simulation and parameter estimation, including the Taiji noise model, the foreground from double white dwarfs (DWDs), and the extragalactic compact binary (ECB) foreground. The methodology for simulating Taiji data and inferring model parameters is described in \Sec{method}. Finally, we summarize our findings and provide concluding remarks in \Sec{conclusion}.

%%%%%%%%%%%%%%%%%%%%%%%%%%%%%%%%%%%%%%%%%%%%%%%%%%%%%%%%%%%%%%%%%%%%%%%%%%%%%%%%%%%%%%%%%%%%%%%%
\section{\label{NEC}SGWB from NEC violation during inflation}

The intermediate NEC violation during inflation is a captivating scenario that has been proposed to explain the distinctive features of the primordial GW background. In this section, we will delve into the intricate details of this model, as outlined by~\cite{Ye:2023tpz}.

The Universe is assumed to begin with an initial phase of slow-roll inflation, characterized by a Hubble parameter $H=H_\text{inf1}$. This stage sets the foundation for the subsequent evolution of the Universe. As the inflationary epoch progresses, a fascinating transition occurs: the Universe enters a second phase of slow-roll inflation, marked by a significantly larger Hubble parameter, $H=H_\text{inf2}$. This transition is not instantaneous; instead, it is punctuated by an intermediate stage of NEC violation, which serves as a bridge between the two slow-roll inflationary phases. Throughout this intricate evolution, a remarkable phenomenon takes place: the comoving Hubble horizon, $a^{-1}H^{-1}$, undergoes a continuous decrease. This reduction in the comoving Hubble horizon has profound implications for the behaviour of perturbation modes. As the Universe experiences accelerated expansion during inflation, perturbation modes gradually exit the horizon, crossing the threshold from the sub-horizon to the super-horizon regime. These modes remain in the super-horizon state until they re-enter the horizon during the ensuing radiation-dominated or matter-dominated era.

The power spectra of perturbation modes exiting the horizon during the first ($k < k_1$) and second ($k > k_2$) stages of slow-roll inflation exhibit a remarkable property: they are nearly scale-invariant. This scale-invariance is a crucial feature, as it aligns with the observations of the cosmic microwave background temperature anisotropies. However, a striking difference emerges between the two stages. The power spectrum amplitude of modes exiting during the second stage is significantly larger than that of the modes exiting during the first stage. This amplification in the power spectrum amplitude is a hallmark of the intermediate NEC violation scenario. The scale-invariance of the tensor power spectrum at large scales has important consequences for the tensor-to-scalar ratio $r$ and the slow-roll parameter in canonical single-field slow-roll inflation. The non-suppressed nature of these quantities ensures their compatibility with observational constraints. Moreover, at small scales, the scale-invariance of both scalar and tensor power spectra plays a crucial role in maintaining the validity of perturbation theory at higher frequencies. By preventing the power spectra from growing to $\mathcal{O}(1)$, the scale-invariance ensures that the perturbation approach remains valid and reliable.

The primordial tensor power spectrum $P_T$ is a fundamental quantity that encapsulates the essential features of the SGWB produced during the early Universe. In this work, we adopt the parametrization proposed by~\cite{Ye:2023tpz}, which elegantly captures the key characteristics of the power spectrum in the presence of an intermediate NEC violation during inflation:
\begin{equation}\label{PT1211}
P_{T} = P_{T,1} + \frac{\pi}{4}(2-n_T) \frac{k}{k_2}|g(k)|^2 P_{T,2},
\end{equation}
where $P_{T,1}$ and $P_{T,2}$ represent the power spectrum amplitudes during the first and second slow-roll inflationary stages, and the auxiliary function $g(k)$ is
\begin{equation} \label{eq:g}
g(k) = H_{\frac{3-n_T}{2}}^{(1)} \left[ \frac{2-n_T}{2} \frac{k}{k_2} \right] \sin \frac{k}{k_2} + H_{\frac{1-n_T}{2}}^{(1)} \left[ \frac{2-n_T}{2} \frac{k}{k_2} \right]\left( \cos \frac{k}{k_2} - \frac{k_2}{k} \sin \frac{k}{k_2} \right) ~.\textbf{}
\end{equation}
Here, $n_T$ denotes the tensor spectral index during the NEC-violating stage and $H_{\nu}^{(1)}$ is the Hankel function of the first kind. 
In the limit of $k \ll k_1$, the primordial tensor power spectrum $P_T$ is approximately equal to $P_{T,1}$, as the second term in Eq.~\eqref{PT1211} becomes negligible compared to $P_{T,1}$. On the other hand, for $k \gg k_1$, the second term in Eq.~\eqref{PT1211} becomes dominant, ensuring that the features around $k \simeq k_2$ are well captured. Although this formulation may sacrifice some accuracy around the first transition scale $k \simeq k_1$, the impact of this compromise on our interests is minimal. 
The primordial tensor power spectrum $P_T$ can be converted to the GW energy spectrum using the relation given by~\cite{Turner:1993vb}:
\begin{equation}
\begin{aligned}
\Omega_{\rm{GW}} &= \frac{k^2}{12 H_0^2}\left[\frac{3\Omega_mj_l(k\tau_0)}{k\tau_0}\sqrt{1.0 + 1.36\frac{k}{k_{\text{eq}}}+2.50\left(\frac{k}{k_{\text{eq}}}\right)^2}\right]^2P_T\\
&\simeq \frac{1}{24}\left(\frac{k}{H_0}\right)^2\left[\frac{3\Omega_m}{(k\tau_0)^2}\sqrt{1.0 + 1.36\frac{k}{k_{\text{eq}}}+2.50\left(\frac{k}{k_{\text{eq}}}\right)^2}\right]^2P_T
\end{aligned}
\end{equation}
where $\tau_{0} = 1.41\times10^{4}$ Mpc, $H_0 = 67.4\, {\rm km/s/Mpc}$ is the Hubble constant~\cite{Planck:2018vyg}, $k_{\text{eq}} = 0.073\,\Omega_{\text{m}} h^{2}$ Mpc$^{-1}$ is the wavenumber at matter-radiation equality, $\Omega_{\rm m}$ is the density fraction of matter today, and the wavenumber $k$ is related to the frequency $f$ by $k = 2\pi f$.
This relation allows us to translate the primordial tensor power spectrum, which is a function of wavenumber $k$, into the observable GW energy spectrum $\Omega_{\rm{GW}}$. The GW energy spectrum represents the fraction of the critical energy density of the Universe that is contained in GWs at a given frequency. As previously discussed, our focus lies on the case where $k_1 \ll k$. In this regime, the first term in Eq. \eqref{PT1211} becomes significantly smaller than the second term, allowing us to safely neglect $P_{T,1}$. This simplification leads to a more manageable parametrization of the theoretical spectrum, which is now fully characterized by three parameters: ${P_{T,2}, \ n_T, \ f_c}$. The transition frequency $f_c$ is related to the characteristic scale $k_2$ by the relation $f_c \equiv 2\pi k_2$.

%%%%%%%%%%%%%%%%%%%%%%%%%%%%%%%%%%%%%%%%%%%%%%%%%%%%%%%%%%%%%%%%%%%%%%%%%%%%%%%%%%%%%%%%%%%%%%%
\begin{figure}[htbp!]
	\centering
	\includegraphics[width=0.8\linewidth]{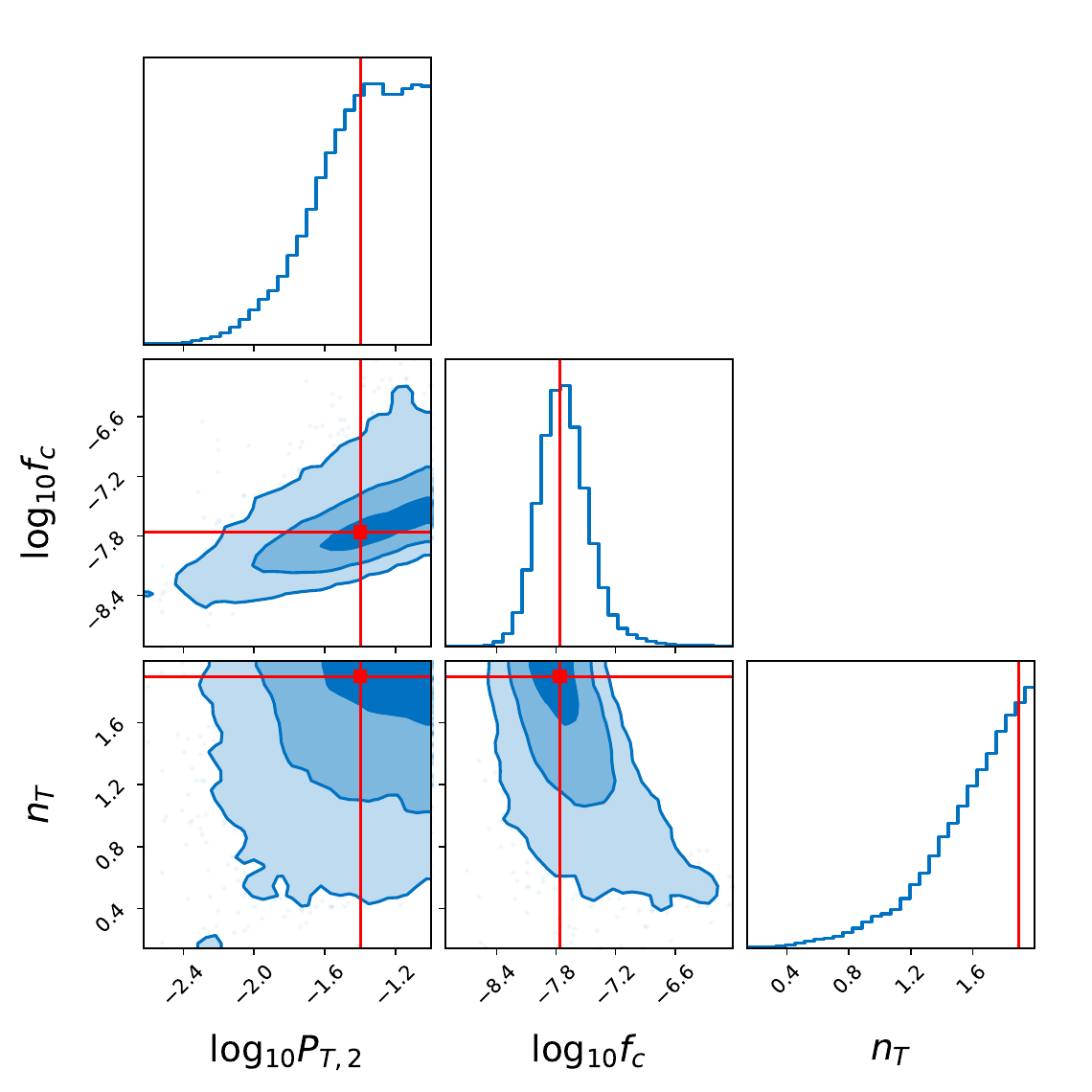}
	\caption{\label{posts_NG}Parameter estimation results derived from the NANOGrav 15-year data set. The red dot represents the selected values used as the basis for the Taiji data injections. The contours in the 2D plots correspond to the $1\sigma$, $2\sigma$, and $3\sigma$ confidence levels in the parameter space.}
\end{figure}
%%%%%%%%%%%%%%%%%%%%%%%%%%%%%%%%%%%%%%%%%%%%%%%%%%%%%%%%%%%%%%%%%%%%%%%%%%%%%%%%%%%%%%%%%%%%%%%%

%%%%%%%%%%%%%%%%%%%%%%%%%%%%%%%%%%%%%%%%%%%%%%%%%%%%%%%%%%%%%%%%%%%%%%%%%%%%%%%%%%%%%%%%%%%%%%%%
\begin{figure}[tbp!]
	\centering
	\includegraphics[width=\linewidth]{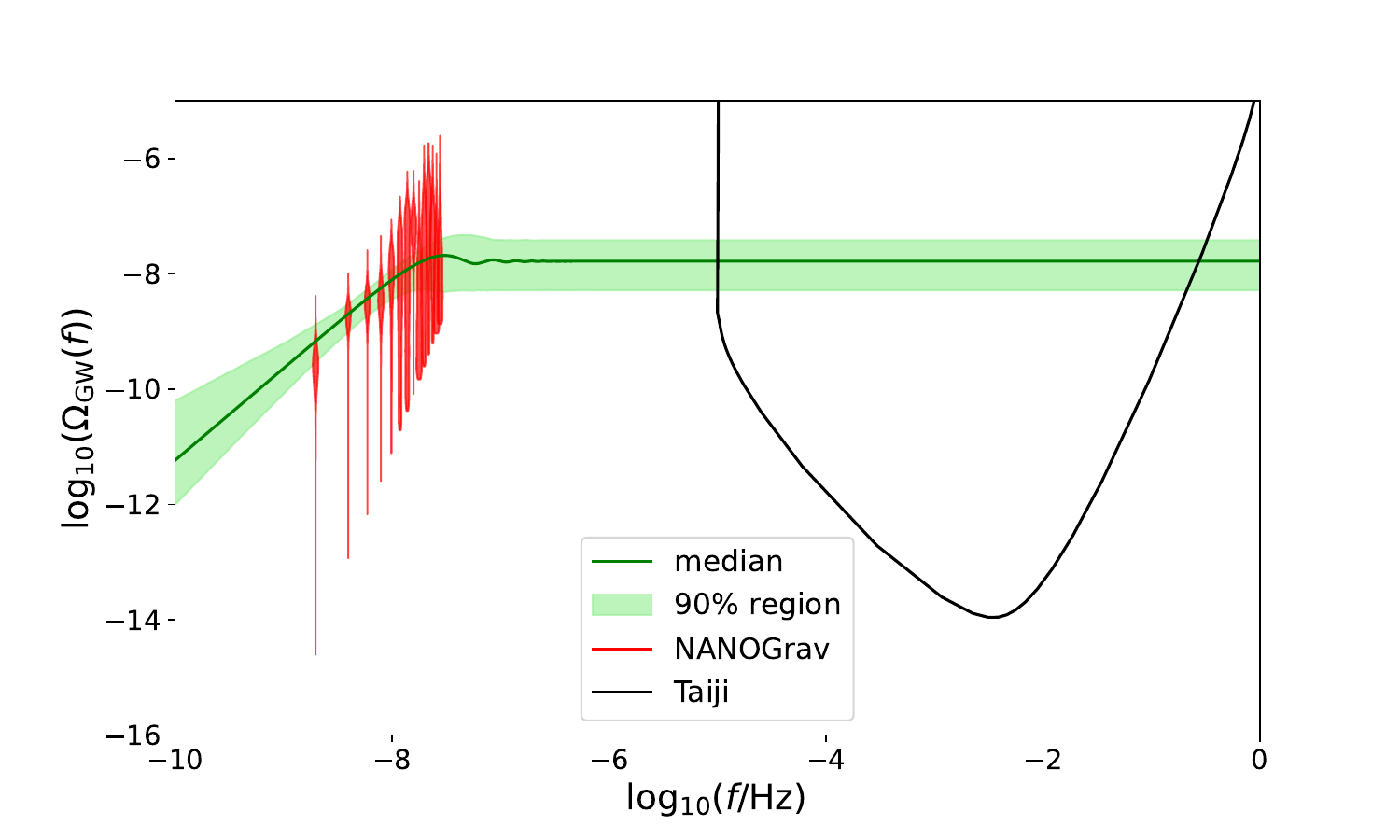}
	\caption{\label{ogw_NEC}Posterior predictive distribution of the energy density spectrum resulting from the NEC violation during inflation (green), along with the free spectrum derived from the NANOGrav 15-year data set (red) and the power-law integrated sensitivity curves for the Taiji detector (black dashed line).}
\end{figure}
%%%%%%%%%%%%%%%%%%%%%%%%%%%%%%%%%%%%%%%%%%%%%%%%%%%%%%%%%%%%%%%%%%%%%%%%%%%%%%%%%%%%%%%%%%%%%%%%

The recently detected PTA signal can potentially be explained by the SGWB produced by primordial NEC violation during inflation~\cite{Ye:2023tpz}. To constrain the model parameters using the NANOGrav 15-year data set, we perform a Bayesian inference following the methodology outlined in~\cite{Bi:2023tib,Liu:2023ymk,Wu:2023hsa,Jin:2023wri,Liu:2023pau}, taking into account the Hellings-Downs correlations. The resulting posterior distributions, shown in \Fig{posts_NG}, yield the following parameter estimates: $P_{T,2} = 4.3^{+4.9}_{-3.0}\times 10^{-2}$, $f_c = 1.9^{+3.3}_{-1.0} \times 10^{-8}\, {\mathrm{Hz}}$, and $n_T = 1.68^{+0.29}_{-0.72}$. These posterior distributions are broadly consistent with those reported in~\cite{Ye:2023tpz}.

We should note that the GW energy density contributes to the effective number of relativistic degrees of freedom, $N_\text{eff}$, at the time of Big Bang Nucleosynthesis (BBN), which can be constrained by observations  \cite{Meerburg:2015zua,Ben-Dayan:2019gll}:
\begin{equation}
N_\text{eff}^\text{GW} = \left( 3.046 + \frac{8}{7} \left( \frac{11}{4} \right) ^{4/3} \right) \frac{1}{12} \int P_T\, \mathrm{d}\ln k.
\end{equation}
Current constraints limit the deviation in $N_\text{eff}$ to $\Delta N_\text{eff} \lesssim 0.4$ at a $95\%$ confidence level \cite{Planck:2018vyg}. To avoid conflicting with the BBN bound, the highly-blue tensor tilt observed in the PTA band must be cut off at a certain wavenumber $k_{cut}=2\pi f_{cut}$ beyond the PTA band. The choice of the cutoff frequency, $f_{cut}$, depends on various assumptions \cite{Meerburg:2015zua,Cabass:2015jwe}. In this study, we consider $f_{cut}=1$ Hz, which produces $\Delta N_\text{eff} \simeq 0.4$ for the injected value, consistent with current constraints. Furthermore, if the power spectrum remains flat up to frequencies between 20 and 76.6 Hz, the energy density of the GWs would violate the upper limit set by the LIGO/Virgo constraint, $\Omega_\text{GW} \lesssim 5.8\times 10^{-9}$, in this frequency band \cite{KAGRA:2021kbb}. Therefore, the cutoff must be chosen to avoid conflicting with these observational results.

\Fig{ogw_NEC} illustrates the posterior predictive distribution for the GW energy density, along with the power-law integrated sensitivity curves of the Taiji detector. The wide frequency range covered by the energy density spectrum suggests that the SGWB generated by NEC violation during inflation could be detectable by Taiji. Motivated by these findings, we will focus on examining the detection capability of the Taiji detector for this specific SGWB signal in the following analysis.

%%%%%%%%%%%%%%%%%%%%%%%%%%%%%%%%%%%%%%%%%%%%%%%%%%%%%%%%%%%%%%%%%%%%%%%%%%%%%%%%%%%%%%%%%%%%%%%%
\section{\label{model}Model components}
In this section, we present the key components of our model for analyzing the sensitivity of the Taiji mission to SGWBs. We consider the instrumental noise characteristics of the Taiji detector, as well as two important astrophysical foregrounds: the DWD foreground and the ECB foreground.

%%%%%%%%%%%%%%%%%%%%%%%%%%%%%%%%%%%%%%%%%%%%%%%%%%%%%%%%%%%%%%%%%%%%%%%%%%%%%%%%%%%%%%%%%%%%%%%%
\subsection{\label{noise}Taiji noise model}
The triangular geometry of the Taiji detector enables the combination of interferometric phase differences using the time delay interferometry (TDI) techniques to suppress laser frequency noise~\cite{Tinto:2001ii,Tinto:2002de}. This process results in the creation of three independent GW measurement channels: the $X, Y, Z$ TDI channels~\cite{Vallisneri:2012np}. Accurately characterizing the noise properties in the Taiji mission is a complex technical challenge, and a detailed analysis of the noise intricacies is beyond the scope of this work. For the purposes of this study, we make several simplifying assumptions regarding the SGWB signal and the instrument noise. We assume that the SGWB signal in the $X, Y, Z$ channels is stationary and uncorrelated with the Taiji instrument noise. Additionally, we model the instrument noise as consisting of test mass acceleration noise and optical path length fluctuation noise, which are assumed to be identical in each spacecraft. Furthermore, we consider the Taiji instruments' arm lengths to be equal, forming an equilateral triangle with arm lengths $L_1=L_2=L_3=L=3 \times 10^9\, \mathrm{m}$. Under these assumptions, the response functions and cross-spectra of the $X, Y, Z$ channel combinations are found to be identical~\cite{Flauger:2020qyi}.
These simplifications allow us to focus on the key aspects of the Taiji detector's performance in detecting the SGWB signal while maintaining a tractable analysis. By assuming stationarity and the absence of correlation between the SGWB signal and the instrument noise, we can treat the two components independently and simplify the statistical analysis. The assumption of identical noise properties in each spacecraft and equal arm lengths further reduces the complexity of the problem, enabling us to derive concise expressions for the response functions and cross-spectra of the TDI channels. While these assumptions may not capture the full complexity of the Taiji mission, they provide a solid foundation for understanding the detector's sensitivity to the SGWB and serve as a starting point for more detailed analyses in the future.

The noise components of the Taiji instrument can be effectively described by two distinct functions. The high-frequency noise, originating from the optical metrology system, is characterized by the power spectral density (PSD) as~\cite{Ren:2023yec}
\begin{equation} 
P_{\mathrm{oms}}(f) = P^2 \times 10^{-24} \frac{1}{\mathrm{Hz}}\left[1+\left(\frac{2\,\mathrm{mHz}}{f}\right)^4\right]\left(\frac{2 \pi f}{c} \mathrm{m}\right)^2,
\end{equation}
where $f$ denotes the frequency, $c$ represents the speed of light, and $P=8$~\cite{Luo:2019zal}. The low-frequency noise, associated with test mass acceleration, is captured by the PSD~\cite{Ren:2023yec} 
\begin{equation}
P_{\mathrm{acc}}(f) = A^2 \times 10^{-30} \frac{1}{\mathrm{Hz}}\left[1+\left(\frac{0.4\, \mathrm{mHz}}{f}\right)^2\right]\left[1+\left(\frac{f}{8\, \mathrm{mHz}}\right)^4\right]\left(\frac{1}{2 \pi f c} \frac{\mathrm{m}}{\mathrm{s}^2} \right)^2,
\end{equation}
where $A=3$~\cite{Luo:2019zal}.
The total PSD for the noise auto-correlation is given by
\begin{equation}\label{auto}
N_{a a}(f, A, P)=16 \sin ^2\left(\frac{f}{f_*}\right)\left\{\left[3+\cos \left(\frac{2f}{f_*}\right)\right] P_{\mathrm{acc}}(f, A)+P_{\mathrm{oms}}(f, P)\right\},
\end{equation}
where $f_*=c / (2 \pi L)$ is a characteristic frequency determined by the arm length $L$. 
The noise cross-spectra are expressed as
\begin{equation}\label{cross}
N_{a b}(f, A, P)=-8 \sin ^2\left(\frac{f}{f_*}\right) \cos \left(\frac{f}{f_*}\right)\left[4 P_{\mathrm{acc}}(f, A)+P_{\mathrm{oms}}(f, P)\right],
\end{equation}
where $a, b \in\{\mathrm{X}, \mathrm{Y}, \mathrm{Z}\}$ and $a \neq b$. Note that the noise covariance matrix is real under these assumptions.

To simplify the analysis, we can utilize linear combinations of the channels. For the space-borne GW instrument, three particular channels are frequently employed: the noise ``orthogonal" channels $A$ and $E$, and the ``null" channel $T$. These channels are defined as follows:
\begin{equation}
\left\{\begin{array}{l}
A=\frac{1}{\sqrt{2}}(Z-X), \\
E=\frac{1}{\sqrt{6}}(X-2 Y+Z), \\
T=\frac{1}{\sqrt{3}}(X+Y+Z) .
\end{array}\right.
\end{equation}
These combinations offer the advantage of noise orthogonality and reduced sensitivity to GWs in the null channel $T$. In the AET basis, the noise spectra can be expressed as
\m
N_{\mathrm{A}} &=&N_{\mathrm{E}}=N_{\mathrm{XX}}-N_{\mathrm{XY}},\\
N_{\mathrm{T}} &=&N_{\mathrm{XX}}+2 N_{\mathrm{XY}}.
\n
By substituting the noise auto- and cross-spectra from \Eq{auto} and \Eq{cross} into these expressions, we obtain the following results for the noise spectra in the AE and T channels:
\begin{equation}
N_{\mathrm{A,E}}=8 \sin ^2\left(\frac{f}{f_*}\right)  \left\{4\left[1+\cos \left(\frac{f}{f_*}\right)+\cos^2\left(\frac{f}{f_*}\right)\right] P_{\mathrm{acc}} +\left[2+\cos \left(\frac{f}{f_*}\right)\right] P_{\mathrm{oms}}\right\},
\end{equation}
and
\begin{equation}
N_{\mathrm{T}}=16 \sin ^2\left(\frac{f}{f_*}\right)  \left\{2\left[1-\cos \left(\frac{f}{f_*}\right)\right]^2 P_{\mathrm{acc}} + \left[1-\cos \left(\frac{f}{f_*}\right)\right] P_{\mathrm{oms}} \right\}. 
\end{equation}
The use of the AET basis and these noise spectra expressions greatly simplifies the analysis of the Taiji instrument's sensitivity to GWs, as it allows for a clear separation of the noise contributions and a straightforward comparison of the noise levels in the different channels.

To facilitate our analysis, we define the equivalent energy spectral density as
\begin{equation}
\Omega_{\alpha}(f)= S_{\alpha}(f) \frac{4 \pi^2 f^3}{3 H_0^2},
\end{equation}
where $\alpha \in \{\mathrm{A, E, T}\}$ denotes the different channel combinations, and $S_{\alpha}$ represents the corresponding noise spectral densities. These noise spectral densities can be expressed as
\m
S_\mathrm{A}(f)&=&S_\mathrm{E}(f)=\frac{N_\mathrm{A}(f)}{\mathcal{R}_\mathrm{A, E}(f)},\\
S_\mathrm{T}(f)&=&\frac{N_\mathrm{T}(f)}{\mathcal{R}_\mathrm{T}(f)},
\n
where $\mathcal{R}_{\alpha}$ denotes the response functions for each channel. Approximate expressions for these response functions have been provided in~\cite{Smith:2019wny}, which take the form
\m
\mathcal{R}_\mathrm{A}^{\mathrm{Fit}}(f)&=&\mathcal{R}_\mathrm{E}^{\mathrm{Fit}}(f)=\frac{9}{20}|W(f)|^2\left[1+\left(\frac{f}{4 f_* / 3}\right)^2\right]^{-1}, \label{RA}\\
\mathcal{R}_\mathrm{T}^{\mathrm{Fit}} &\simeq& \frac{1}{4032}\left(\frac{f}{f_*}\right)^6|W(f)|^2\left[1+\frac{5}{16128}\left(\frac{f}{f_*}\right)^8\right]^{-1},\label{RB}
\n
where $W(f)=1-e^{-2 i f / f_*}$.
While these approximate expressions can be useful, in this study, we employ the more accurate analytical expressions for the response functions that were recently derived in Ref.~\cite{Wang:2021owg}. These analytical expressions provide a more precise description of the detector's response and are essential for the accurate characterization of the GW signal.
{While this choice may have some impact on the quantitative results, we expect the overall conclusions of our analysis to remain robust, as the differences between the ORFs from Ref.~\cite{Wang:2021owg} and Eqs.~\eqref{RA} and \eqref{RB} are relatively small.}

%%%%%%%%%%%%%%%%%%%%%%%%%%%%%%%%%%%%%%%%%%%%%%%%%%%%%%%%%%%%%%%%%%%%%%%%%%%%%%%%%%%%%%%%%%%%%%%%
\subsection{Double white dwarf foreground}
Astrophysical population models~\cite{Korol:2020lpq,Korol:2021pun} predict the existence of tens of millions of DWD binaries within our Milky Way galaxy. These DWD binaries are expected to emit GWs simultaneously within a frequency range spanning from $10^{-5}\, \mathrm{Hz}$ to $0.1\, \mathrm{Hz}$~\cite{Karnesis:2021tsh}. While a small fraction of these DWD binaries are individually resolvable, the vast majority remain unresolved and collectively contribute to a stochastic ``galactic foreground" or ``confusion noise" for the Taiji mission~\cite{Liu:2023qap}.
The DWD foreground can be approximated by a polynomial function $S_c(f)$ in the logarithmic scale as~\cite{Liu:2023qap}
\begin{equation}\label{SDWD1}
S_{\mathrm{DWD}}(f)=\exp \left(\sum_{i=0}^5 a_i\left(\log \left(\frac{f}{\mathrm{mHz}}\right)\right)^i\right) \mathrm{Hz}^{-1} .
\end{equation}
This approximation is valid for the frequency range $0.1\, \mathrm{mHz}<f<10\, \mathrm{mHz}$. The values of the parameters $a_i$ for a $4$-year observation are $a_0 =-85.5448$, $a_1=-3.23671$, $a_2=-1.64187$, $a_3=-1.14711$, $a_4=0.0325887$, $a_5=0.187854$. The corresponding dimensionless energy spectral density is given by
\begin{equation}
\Omega_{\mathrm{DWD}}(f)=S_{\mathrm{DWD}}(f) \frac{4 \pi^2 f^3}{3 H_0^2}.
\end{equation} 

In this work, we employ a broken power law to model the dimensionless energy spectral density $\Omega_{\mathrm{DWD}}$ of the DWD foreground. The expression for $\Omega_{\mathrm{DWD}}(f)$ is given by
\begin{equation}\label{SDWD2}
\Omega_{\mathrm{DWD}}(f)=\frac{A_1\left(f / f_*\right)^{\alpha_1}}{1+A_2\left(f / f_*\right)^{\alpha_2}},
\end{equation}
with the following parameter values: $A_1=3.98\times 10^{-16}$, $A_2 = 4.79\times 10^{-7}$, $\alpha_1= -5.7$, and $\alpha_2=-6.2$~\cite{Chen:2023zkb}.
The broken power-law spectral shape of the DWD foreground arises from the physical constraint imposed by the finite radii of the two white dwarfs in each binary system. As the frequency increases, the number of DWDs capable of emitting at those frequencies decreases, leading to a change in spectral behaviour. {We have chosen to use the broken power law in \Eq{SDWD2} with four parameters to represent the DWD foreground, as it provides a simpler and more compact description compared to \Eq{SDWD1} with six parameters. In the frequency range of interest for our analysis, these two parameterizations exhibit good agreement. Consequently, we anticipate minimal impact on the recovery of the SGWB signal, particularly when the SGWB signal is significantly stronger than the DWD foreground, as illustrated in \Fig{data}.}
% This characteristic spectral shape provides valuable information about the properties and distribution of DWD binaries in our galaxy, and its accurate modelling is crucial for disentangling the DWD foreground from other GW signals of interest in the Taiji mission data.

%%%%%%%%%%%%%%%%%%%%%%%%%%%%%%%%%%%%%%%%%%%%%%%%%%%%%%%%%%%%%%%%%%%%%%%%%%%%%%%%%%%%%%%%%%%%%%%%
\subsection{Extragalactic compact binary foreground}
The existence of a background originating from compact binaries, comprising black holes and neutron stars, in distant galaxies is predicted by current models~\cite{Chen:2018rzo}. Nevertheless, ground-based GW observatories have not yet succeeded in detecting this ECB foreground.
In this study, we model the energy spectral density of the ECB foreground using a power-law approximation:
\begin{equation}
\Omega_{\mathrm{ECB}}(f) = A_{\mathrm{ECB}}\left(\frac{f}{f_{\mathrm{ref}}}\right)^{\alpha_{\mathrm{ECB}}},
\end{equation}
where $f_{\mathrm{ref}} = 25 \mathrm{Hz}$ is the reference frequency, and $\alpha_{\mathrm{ECB}} = \frac{2}{3}$ represents the power-law index. The amplitude $A_{\mathrm{ECB}}$ is estimated to be $1.8 \times 10^{-9}$, based on the findings presented in~\cite{Chen:2018rzo}.
The inclusion of the ECB foreground is crucial for accurately assessing the mission's ability to detect and characterize other sources of SGWBs, such as those originating from the early Universe or more exotic phenomena. As the sensitivity of GW detectors continues to improve, the detection and precise characterization of the ECB foreground will become increasingly important for advancing our understanding of the population and evolution of compact binary systems throughout the cosmos.
%%%%%%%%%%%%%%%%%%%%%%%%%%%%%%%%%%%%%%%%%%%%%%%%%%%%%%%%%%%%%%%%%%%%%%%%%%%%%%%%%%%%%%%%%%%%%%%%
\section{\label{method}Methodology}

%%%%%%%%%%%%%%%%%%%%%%%%%%%%%%%%%%%%%%%%%%%%%%%%%%%%%%%%%%%%%%%%%%%%%%%%%%%%%%%%%%%%%%%%%%%%%%%%

%%%%%%%%%%%%%%%%%%%%%%%%%%%%%%%%%%%%%%%%%%%%%%%%%%%%%%%%%%%%%%%%%%%%%%%%%%%%%%%%%%%%%%%%%%%%%%%
\begin{figure}[htbp!]
	\centering
	\includegraphics[width=\linewidth]{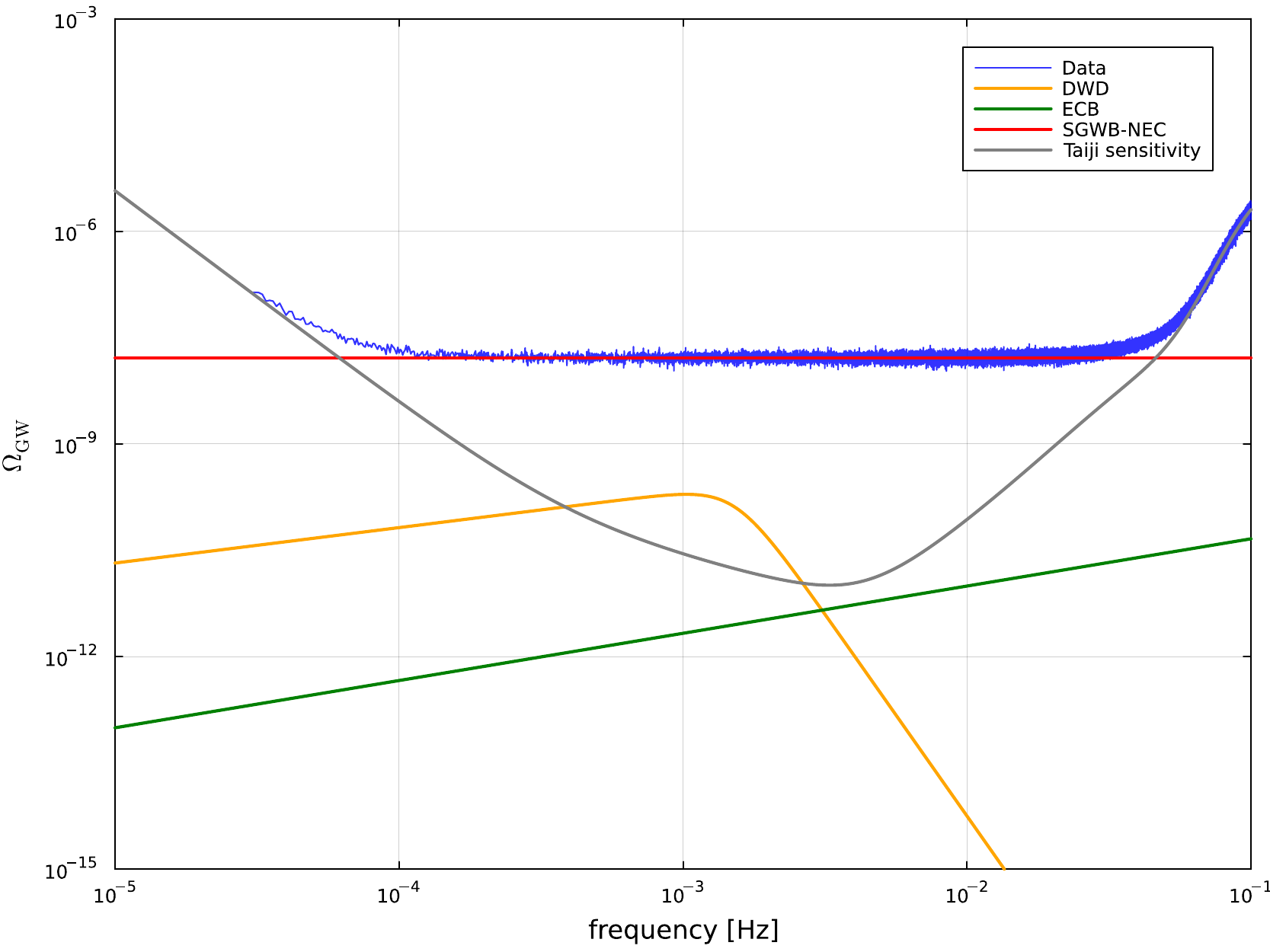}
	\caption{\label{data}Simulated A-channel data for the Taiji detector in the frequency domain (blue), along with the DWD foreground (orange), ECB foreground (green), and the SGWB produced by the NEC violation (red). The Taiji sensitivity curve is also displayed, represented by the dimensionless energy spectral density $\Omega_{\mathrm{GW}}(f)$.}
\end{figure}
%%%%%%%%%%%%%%%%%%%%%%%%%%%%%%%%%%%%%%%%%%%%%%%%%%%%%%%%%%%%%%%%%%%%%%%%%%%%%%%%%%%%%%%%%%%%%%%%
In this study, we simulate the Taiji data, closely adhering to the procedures outlined in Refs.~\cite{Caprini:2019pxz,Flauger:2020qyi}. We consider a scenario where data is collected over a full mission duration of $4$ years, with an estimated efficiency of $75\%$, resulting in an effective observation time of $3$ years. The TDI data is segmented into approximately $N_\mathrm{c} = 94$ chunks, each spanning $11.5$ days~\cite{Caprini:2019pxz,Flauger:2020qyi}. The Taiji mission's frequency range of interest extends from  $3 \times 10^{-5}\, \mathrm{Hz}$ to $5 \times 10^{-1}\, \mathrm{Hz}$, with a frequency resolution of approximately $10^{-6}\, \mathrm{Hz}$. This results in a total of roughly $5 \times 10^7$ data points for the entire simulation.

To analyze the data stream, we perform a Fourier transform, considering only positive frequencies due to the real nature of the time stream. The time data stream can be expressed as
\begin{equation}
d(t) = \sum_{f=f_{\min}}^{f_{\max}} \left[d(f) e^{-2 \pi i f t} + d^*(f) e^{2 \pi i f t}\right],
\end{equation}
where $d(f)$ represents the Fourier amplitudes at frequency $f$, and $d^{*}(f)$ denotes the complex conjugate of $d(f)$. 
Assuming the SGWB and the noise are both stationary, namely $\langle d(t) d(t^{\prime}\rangle=f(t-t^{\prime})$ and $\langle d(t)\rangle = 0$, the ensemble averages of the Fourier coefficients satisfy
\begin{equation}
\left\langle d(f) d\left(f^{\prime}\right)\right\rangle=0 \quad \text { and } \quad\left\langle d(f) d^*\left(f^{\prime}\right)\right\rangle=D(f) \delta_{f f^{\prime}},
\end{equation}
indicating that the Fourier coefficients at different frequencies are uncorrelated, and the correlation strength at the same frequency is determined by the PSD, $D(f)$.

\begin{table}
\centering
\begin{tabular}{c|c|c|c}
\hline\hline
Parameter & Prior & Injected value  & Recovered value \\ 
\hline
$A$ & $\mU(2.95, 3.05)$ & $3$  & $3.00^{+0.01}_{-0.01}$ \\ 
$P$ & $\mU(7.99, 8.01)$ & $8$  & $8.000^{+0.002}_{-0.002}$ \\ 
$\log_{10} A_1$ & $\mU(-17, -14)$ & $-15.4$  & $-15.3^{+0.7}_{-0.8}$ \\ 
$\alpha_1$ & $\mU(-7, -4)$ & $-5.7$  & $-6.0^{+1.1}_{-0.9}$ \\ 
$\log_{10} A_2$ & $\mU(-8, -5)$  & $-6.32$  & $-6.13^{+0.53}_{-0.89}$ \\ 
$\alpha_2$ & $\mU(-9, -5)$ & $-6.2$  &$-6.4^{+0.8}_{-0.9}$ \\ 
$\log_{10} A_{\mathrm{ECB}}$ & $\mU(-10, -7)$ & $-8.74$  & $-9.13^{+0.91}_{-0.42}$ \\ 
$\alpha_{\mathrm{ECB}}$ & $\mU(0.2, 1)$ & $2/3$  & $0.67^{+0.29}_{-0.38}$ \\ 
$\log_{10} P_{T,2}$ & $\mU(-1.41, -1.39)$ & $-1.4$  & $-1.4000^{+0.0002}_{-0.0003}$ \\ 
$\log_{10} (f_c /\mathrm{Hz})$ & $\mU(-9, -6)$ & $-7.76$  & $-7.47^{+1.30}_{-1.37}$ \\ 
$n_T$ & $\mU(1, 2)$ & $1.9$  & $1.5^{+0.4}_{-0.5}$ \\ 
\hline
\end{tabular}
\caption{\label{tab:priors}Summary of the model parameters and their associated priors used in the Bayesian inference analysis. The injected values used in the simulation and the recovered values obtained from the parameter estimation are provided. The recovered values are presented as the median along with the $90\%$ equal-tail credible intervals. $\mathcal{U}$ denotes the uniform distribution.}
\end{table}

To simulate a signal, we employ Gaussian distributions to generate random variables that correspond to the real and imaginary components of the Fourier coefficients. These variables are drawn from distributions with a variance of $D(f) / 2$. By modelling both the signal and noise as Gaussian processes, we assert that their statistical properties can be completely described by their respective power spectra, denoted by $\Omega_{\mathrm{GW}}(f)$ for the GW signal and $\Omega_{\mathrm{A,E,T}}(f)$ for the detector noise in different channels.
Specifically, at each discrete frequency $f_i$, we compute the signal power, $S_i$, and noise power, $N_i$, as
\m
S_i & =&\left|\frac{G_{i 1}\left(0, \sqrt{\Omega_{\mathrm{GW}}\left(f_i\right)}\right)+i G_{i 2}\left(0, \sqrt{\Omega_{\mathrm{GW}}\left(f_i\right)}\right)}{\sqrt{2}}\right|^2, \\
N_i & =&\left|\frac{G_{i 3}\left(0, \sqrt{\Omega_{\mathrm{A,E,T}}\left(f_i\right)}\right)+i G_{i 4}\left(0, \sqrt{\Omega_{\mathrm{A,E,T}}\left(f_i\right)}\right)}{\sqrt{2}}\right|^2 .
\n
Here, $G_{i 1}(M, \sigma), \ldots, G_{i 4}(M, \sigma)$ represent Gaussian-distributed real numbers with mean $M$ and standard deviation $\sigma$. These are the stochastic realizations of the real and imaginary parts of the Fourier coefficients for both signal and noise.
The total power at each frequency, $D_i$, is obtained by summing the corresponding signal and noise powers, which we assume to be independent:
\begin{equation}
D_i = S_i + N_i.
\end{equation}
For each frequency bin indexed by $i$, we generate a set of $N_\mathrm{c}$ data points, ${D_{i1}, D_{i2}, \ldots, D_{iN_\mathrm{c}}}$, and compute their mean, $\bar{D}_i$. 
{It is worth noting that the independent generation of the noise realizations and the SGWB signal realization implicitly assumes that the SGWB signal is uncorrelated with the detector noises, as mentioned in Section \ref{noise}. This assumption is reflected in the simulated data, where the SGWB signal and detector noises are added together without any correlation between them.}
The simulated data, depicted in \Fig{data}, is generated using parameter values listed in \Table{tab:priors}.

%%%%%%%%%%%%%%%%%%%%%%%%%%%%%%%%%%%%%%%%%%%%%%%%%%%%%%%%%%%%%%%%%%%%%%%%%%%%%%%%%%%%%%%%%%%%%%%
\begin{figure}[tbp!]
	\centering
	\includegraphics[width=\linewidth]{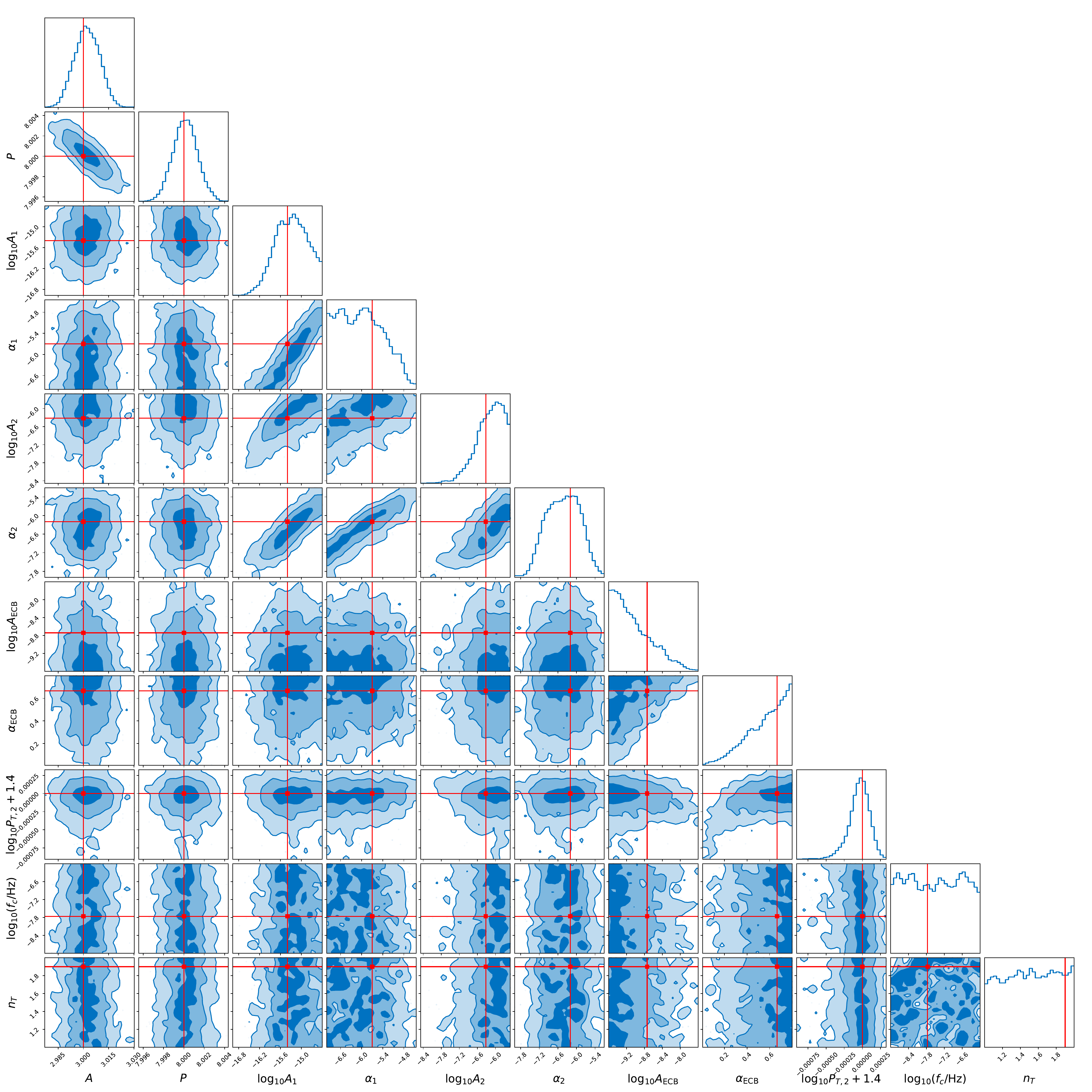}
	\caption{\label{posts_Taiji}Posterior distributions of the model parameters for the NEC violation model, as inferred by the Taiji detector. The red points represent the injected values for each parameter, serving as a reference to assess the accuracy of the parameter estimation. The contours in the two-dimensional parameter space correspond to the $1\sigma$, $2\sigma$, and $3\sigma$ confidence levels.}
\end{figure}

%%%%%%%%%%%%%%%%%%%%%%%%%%%%%%%%%%%%%%%%%%%%%%%%%%%%%%%%%%%%%%%%%%%%%%%%%%%%%%%%%%%%%%%%%%%%%%%
\begin{figure}[tbp!]
	\centering
	\includegraphics[width=\linewidth]{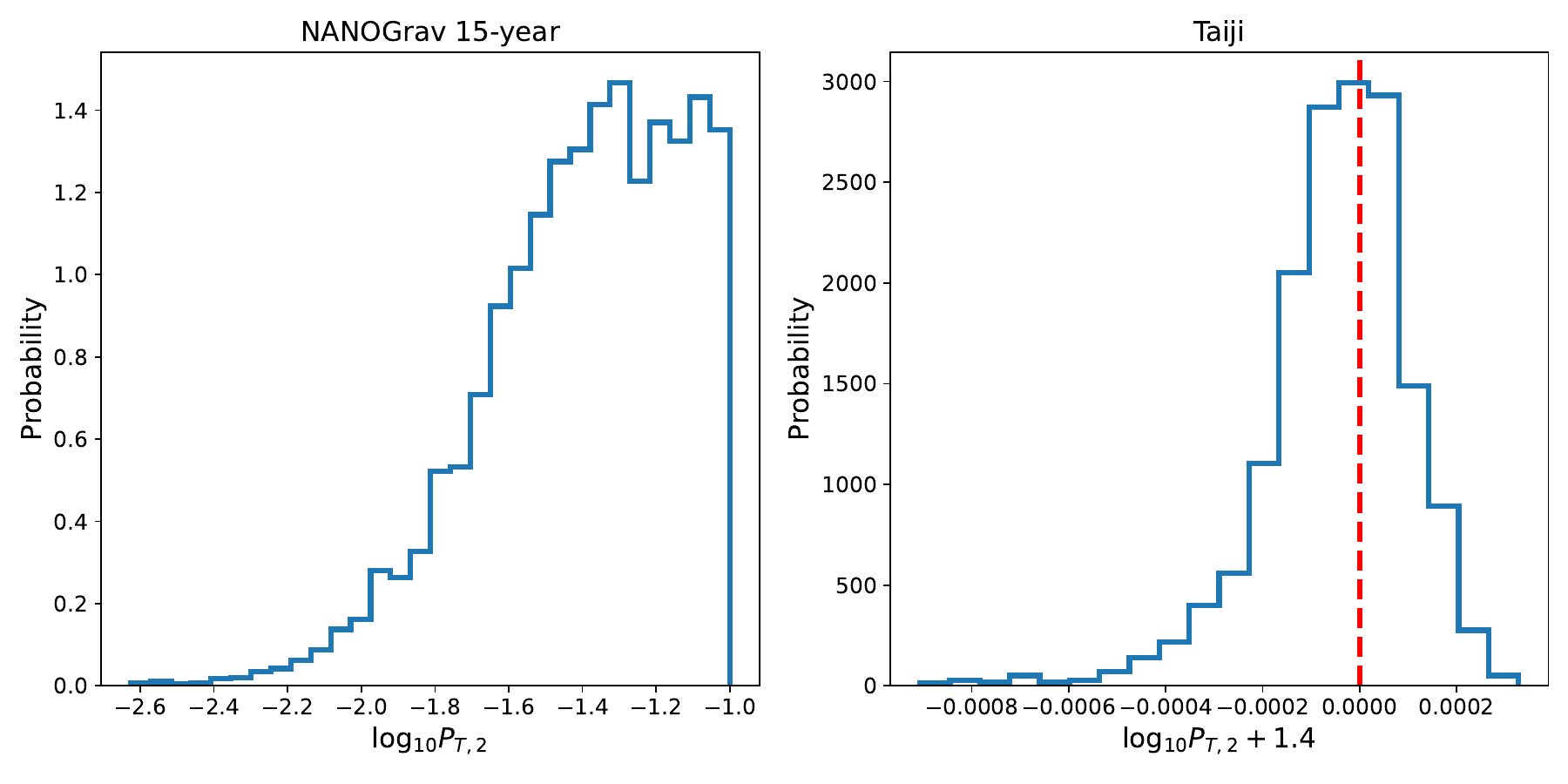}
	\caption{\label{compare}Comparison of the posterior distributions for the tensor power spectrum amplitude during the second inflationary stage, $P_{T,2}$, derived from the NANOGrav 15-year data set (left panel) and the simulated Taiji mission data (right panel). The plot demonstrates the substantial improvement in the precision of the $P_{T,2}$ measurement achieved by Taiji, as evidenced by the significantly narrower posterior distribution compared to that obtained from the PTA data.}
\end{figure}

To streamline the computational demands associated with high-frequency data, we employ a rebinning strategy, termed ``coarse graining", for frequencies from $f=10^{-3}\, \mathrm{Hz}$ to $f_{\max}=0.5\, \mathrm{Hz}$. The rebinning process consolidates the frequency range into $1000$ equally spaced logarithmic intervals. Frequencies in the lower spectrum, specifically from $f_{\min}=3 \times 10^{-5}\,\mathrm{Hz}$ to $f=10^{-3}\,\mathrm{Hz}$, are preserved in their original form. The modified data set thus comprises $1971$ bins per segment, optimizing computational efficiency without compromising data integrity.
The rebinned data within this specified frequency domain are redefined as
\m
f_k &\equiv& \sum_{j \in \operatorname{bin} k} w_j f_j, \\
\bar{D}_k &\equiv& \sum_{j \in \operatorname{bin} k} w_j \bar{D}_j,
\n
where $w_j$ are the weighting factors given by
\begin{equation}
w_j =\frac{\mathcal{D}^\mathrm{th}(f_j, \vec{\theta}, \vec{n})^{-1}}{\sum_{l \in \mathrm{bin} k}\mathcal{D}^\mathrm{th}(f_l, \vec{\theta}, \vec{n})^{-1}},
\end{equation}
for each frequency $j$ within the aggregated interval $k$. The theoretical data model, $\mathcal{D}^\mathrm{th}(f, \vec{\theta}, \vec{n})$, is defined by the sum of the GW spectrum, $\Omega_\mathrm{GW}(f, \vec{\theta})$, and detector noise, $\Omega_{\alpha} (f, \vec{n})$. Here, $\vec{n} = \{A, P\}$ represents the noise parameters, while the signal parameters are encapsulated by $\vec{\theta} = \{A_1, \alpha_1, A_2, \alpha_2, A_\mathrm{ECB}, \alpha_\mathrm{ECB}, n_T, f_c, P_{T,2}\}$.
For practical analysis, we use the AET basis, which simplifies the covariance matrix to a diagonal form, facilitating more straightforward computations.

The likelihood function is computed as a composition of the Gaussian part and the log-normal part~\cite{Flauger:2020qyi}. The log-likelihood combination is mathematically represented as
\begin{equation}
\ln \mathcal{L}=\frac{1}{3} \ln \mathcal{L}_\mathrm{G}+\frac{2}{3} \ln \mathcal{L}_\mathrm{LN}.
\end{equation}
The Gaussian component of the likelihood, $\ln \mathcal{L}_\mathrm{G}$, which scrutinizes the residuals between the theoretical model and observed data, is given by
\begin{equation}
\ln \mathcal{L}_\mathrm{G}(D | \vec{\theta}, \vec{n})=-\frac{N_{\mathrm{c}}}{2} \sum_{\alpha} \sum_k n_{\alpha}^{(k)}\left[\frac{\mathcal{D}_{\alpha}^\mathrm{t h}\left(f_{\alpha}^{(k)}, \vec{\theta}, \vec{n}\right)-\mathcal{D}_{\alpha}^{(k)}}{\mathcal{D}_{\alpha}^\mathrm{t h}\left(f_{\alpha}^{(k)}, \vec{\theta}, \vec{n}\right)}\right]^2,
\end{equation}
where the sums iterate over the AET detector channels, denoted by $\alpha$, and the frequency bins, indexed by $k$.
The log-normal component, $\ln \mathcal{L}_\mathrm{LN}$, assesses the logarithmic deviations and is formalized as
\begin{equation}
\ln \mathcal{L}_\mathrm{L N}(D | \vec{\theta}, \vec{n})=-\frac{N_{\mathrm{c}}}{2} \sum_{\alpha} \sum_k n_{\alpha}^{(k)} \ln ^2\left[\frac{\mathcal{D}_{\alpha}^\mathrm{t h}\left(f_{\alpha}^{(k)}, \vec{\theta}, \vec{n}\right)}{\mathcal{D}_{\alpha}^{(k)}}\right].
\end{equation}

%%%%%%%%%%%%%%%%%%%%%%%%%%%%%%%%%%%%%%%%%%%%%%%%%%%%%%%%%%%%%%%%%%%%%%%%%%%%%%%%%%%%%%%%%%%%%%%%
\section{\label{conclusion}Result and discussion}

We employ the \texttt{dynesty}~\cite{Speagle:2019ivv} sampler, accessed through the \texttt{Bilby} package~\cite{Ashton:2018jfp,Romero-Shaw:2020owr}, to thoroughly investigate the parameter space. \Table{tab:priors} provides a comprehensive summary of the model parameters and their corresponding priors. \Fig{posts_Taiji} presents the posterior distributions for the parameters, obtained using data from the Taiji detector. The injected values, denoted in red, are overlaid on these distributions, showing that the parameters can be well recovered within a $2\sigma$ confidence interval. For ease of comparison, the median values and $90\%$ credible intervals are concisely tabulated in \Table{tab:priors}. Our analysis reveals that the noise parameters, denoted by $A$ and $P$, can be precisely determined with relative uncertainties of $0.3\%$ and $0.03\%$, respectively. In contrast, the parameters associated with the ECB and the DWD foregrounds exhibit larger uncertainties due to their more subtle signatures compared to the prominent SGWB from NEC violation. Remarkably, the tensor power spectrum amplitude during the second inflationary stage can be determined with an impressive relative uncertainty of $0.02\%$ with a result of $\log_{10} P_{T,2} = -1.4000^{+0.0002}_{-0.0003}$. This level of precision represents a substantial improvement, more than a factor of $1000$, compared to what was achievable with the NANOGrav 15-year data set. This highlights the exceptional sensitivity of the Taiji detector in refining the measurements of these crucial parameters. To effectively illustrate the progress made, we directly compare the posterior distributions obtained from the NANOGrav 15-year data set with those derived using the Taiji detector, as depicted in \Fig{compare}. This side-by-side comparison demonstrates the remarkable enhancements in parameter estimation achieved with Taiji, emphasizing its potential to deepen our understanding of NEC violation.

In this paper, we have explored the potential of the Taiji space-based GW detector to detect an SGWB originating from the violation of the NEC during inflation. We posited that the stochastic signal observed by PTAs, such as NANOGrav, stems from NEC violation during the inflationary epoch and leveraged the NANOGrav 15-year data set to constrain the model parameters. By conducting parameter estimation on simulated Taiji data, we quantified the uncertainties in determining the model parameters associated with the GW signal from NEC violation during inflation. Our results demonstrate a remarkable improvement in precision compared to the NANOGrav 15-year data set, surpassing it by more than three orders of magnitude. This significant enhancement highlights the superior measurement capabilities of the Taiji detector. Consequently, space-based GW detectors like Taiji will play a pivotal role in verifying the stochastic signal detected by PTAs and serve as invaluable tools for further validation of NEC violation signatures. Moreover, our study underscores the importance of accurately characterizing various components of the model~\cite{Boileau:2021gbr,Chen:2023zkb}, including the Taiji noise model and foreground contributions from DWDs and ECBs. 

{The simulated Taiji data presented in Fig. 4 demonstrates that the detector has limited sensitivity to the transition frequency $f_c$ and the tensor spectral index $n_T$ during the NEC-violating stage. This can be attributed to the fact that the step-like feature in the primordial tensor power spectrum $P_T$, which is a characteristic signature of the NEC violation, lies outside the frequency range probed by Taiji. Consequently, the signal tested in this study effectively resembles a scale-invariant tensor power spectrum within the Taiji frequency band. It is important to note that the primary role of NEC violation in this scenario is to allow for a high primordial tensor amplitude $P_T$ that is detectable by space-based interferometers like Taiji without contradicting the constraints from CMB observations. The NEC violation enables the generation of a significant tensor amplitude at small scales while remaining consistent with the CMB bounds on the tensor-to-scalar ratio at large scales.}

{This finding highlights the importance of exploring alternative models or scenarios where the features induced by NEC violation may fall within the sensitivity range of space-based GW detectors. Such investigations could provide valuable insights into the nature of NEC violation and its observational signatures. Future studies could consider models with different transition scales or more complex dynamics during the NEC-violating phase, which may yield distinct features in the tensor power spectrum that are accessible to space-based interferometers. Moreover, the results presented in this work underscore the complementarity between CMB measurements and direct GW observations in constraining early Universe physics. While CMB observations provide stringent bounds on the tensor amplitude at large scales, space-based interferometers like Taiji offer a unique probe of the primordial tensor spectrum at smaller scales. The combination of these observational techniques can lead to a more comprehensive understanding of the inflationary paradigm and the role of NEC violation in shaping the early Universe.}

In conclusion, the prospects for Taiji to detect an SGWB from NEC violation during inflation appear promising, and our study provides insights into the parameter estimation and detection capabilities of Taiji for NEC violation during inflation. Future observations and analyses from Taiji and other space-based GW detectors will continue to illuminate the origin and properties of the SGWB, contributing to our understanding of the early Universe and fundamental physics. 
%%%%%%%%%%%%%%%%%%%%%%%%%%%%%%%%%%%%%%%%%%%%%%%%%%%%%%%%%%%%%%%%%%%%%%%%%%%%%%%%%%%%%%%%%%%%%%%%
\section*{Acknowledgments}
ZCC is supported by the National Natural Science Foundation of China under Grant No.~12405056 and the innovative research group of Hunan Province under Grant No.~2024JJ1006.
LL is supported by the National Natural Science Foundation of China (Grant No. 12433001, No. 12247112 and No. 12247176) and the China Postdoctoral Science Foundation Fellowship No. 2023M730300.

\bibliography{ref}

\begin{thebibliography}{108}%
\makeatletter
\providecommand \@ifxundefined [1]{%
 \@ifx{#1\undefined}
}%
\providecommand \@ifnum [1]{%
 \ifnum #1\expandafter \@firstoftwo
 \else \expandafter \@secondoftwo
 \fi
}%
\providecommand \@ifx [1]{%
 \ifx #1\expandafter \@firstoftwo
 \else \expandafter \@secondoftwo
 \fi
}%
\providecommand \natexlab [1]{#1}%
\providecommand \enquote  [1]{``#1''}%
\providecommand \bibnamefont  [1]{#1}%
\providecommand \bibfnamefont [1]{#1}%
\providecommand \citenamefont [1]{#1}%
\providecommand \href@noop [0]{\@secondoftwo}%
\providecommand \href [0]{\begingroup \@sanitize@url \@href}%
\providecommand \@href[1]{\@@startlink{#1}\@@href}%
\providecommand \@@href[1]{\endgroup#1\@@endlink}%
\providecommand \@sanitize@url [0]{\catcode `\\12\catcode `\$12\catcode `\&12\catcode `\#12\catcode `\^12\catcode `\_12\catcode `\%12\relax}%
\providecommand \@@startlink[1]{}%
\providecommand \@@endlink[0]{}%
\providecommand \url  [0]{\begingroup\@sanitize@url \@url }%
\providecommand \@url [1]{\endgroup\@href {#1}{\urlprefix }}%
\providecommand \urlprefix  [0]{URL }%
\providecommand \Eprint [0]{\href }%
\providecommand \doibase [0]{http://dx.doi.org/}%
\providecommand \selectlanguage [0]{\@gobble}%
\providecommand \bibinfo  [0]{\@secondoftwo}%
\providecommand \bibfield  [0]{\@secondoftwo}%
\providecommand \translation [1]{[#1]}%
\providecommand \BibitemOpen [0]{}%
\providecommand \bibitemStop [0]{}%
\providecommand \bibitemNoStop [0]{.\EOS\space}%
\providecommand \EOS [0]{\spacefactor3000\relax}%
\providecommand \BibitemShut  [1]{\csname bibitem#1\endcsname}%
\let\auto@bib@innerbib\@empty
%</preamble>
\bibitem [{\citenamefont {Abbott}\ \emph {et~al.}(2016)\citenamefont {Abbott} \emph {et~al.}}]{LIGOScientific:2016aoc}%
  \BibitemOpen
  \bibfield  {author} {\bibinfo {author} {\bibfnamefont {B.~P.}\ \bibnamefont {Abbott}} \emph {et~al.} (\bibinfo {collaboration} {LIGO Scientific, Virgo}),\ }\bibfield  {title} {\enquote {\bibinfo {title} {{Observation of Gravitational Waves from a Binary Black Hole Merger}},}\ }\href {\doibase 10.1103/PhysRevLett.116.061102} {\bibfield  {journal} {\bibinfo  {journal} {Phys. Rev. Lett.}\ }\textbf {\bibinfo {volume} {116}},\ \bibinfo {pages} {061102} (\bibinfo {year} {2016})},\ \Eprint {http://arxiv.org/abs/1602.03837} {arXiv:1602.03837 [gr-qc]} \BibitemShut {NoStop}%
\bibitem [{\citenamefont {Abbott}\ \emph {et~al.}(2019{\natexlab{a}})\citenamefont {Abbott} \emph {et~al.}}]{LIGOScientific:2019fpa}%
  \BibitemOpen
  \bibfield  {author} {\bibinfo {author} {\bibfnamefont {B.~P.}\ \bibnamefont {Abbott}} \emph {et~al.} (\bibinfo {collaboration} {LIGO Scientific, Virgo}),\ }\bibfield  {title} {\enquote {\bibinfo {title} {{Tests of General Relativity with the Binary Black Hole Signals from the LIGO-Virgo Catalog GWTC-1}},}\ }\href {\doibase 10.1103/PhysRevD.100.104036} {\bibfield  {journal} {\bibinfo  {journal} {Phys. Rev. D}\ }\textbf {\bibinfo {volume} {100}},\ \bibinfo {pages} {104036} (\bibinfo {year} {2019}{\natexlab{a}})},\ \Eprint {http://arxiv.org/abs/1903.04467} {arXiv:1903.04467 [gr-qc]} \BibitemShut {NoStop}%
\bibitem [{\citenamefont {Abbott}\ \emph {et~al.}(2021{\natexlab{a}})\citenamefont {Abbott} \emph {et~al.}}]{LIGOScientific:2020tif}%
  \BibitemOpen
  \bibfield  {author} {\bibinfo {author} {\bibfnamefont {R.}~\bibnamefont {Abbott}} \emph {et~al.} (\bibinfo {collaboration} {LIGO Scientific, Virgo}),\ }\bibfield  {title} {\enquote {\bibinfo {title} {{Tests of general relativity with binary black holes from the second LIGO-Virgo gravitational-wave transient catalog}},}\ }\href {\doibase 10.1103/PhysRevD.103.122002} {\bibfield  {journal} {\bibinfo  {journal} {Phys. Rev. D}\ }\textbf {\bibinfo {volume} {103}},\ \bibinfo {pages} {122002} (\bibinfo {year} {2021}{\natexlab{a}})},\ \Eprint {http://arxiv.org/abs/2010.14529} {arXiv:2010.14529 [gr-qc]} \BibitemShut {NoStop}%
\bibitem [{\citenamefont {Abbott}\ \emph {et~al.}(2021{\natexlab{b}})\citenamefont {Abbott} \emph {et~al.}}]{LIGOScientific:2021sio}%
  \BibitemOpen
  \bibfield  {author} {\bibinfo {author} {\bibfnamefont {R.}~\bibnamefont {Abbott}} \emph {et~al.} (\bibinfo {collaboration} {LIGO Scientific, VIRGO, KAGRA}),\ }\bibfield  {title} {\enquote {\bibinfo {title} {{Tests of General Relativity with GWTC-3}},}\ }\href@noop {} {\  (\bibinfo {year} {2021}{\natexlab{b}})},\ \Eprint {http://arxiv.org/abs/2112.06861} {arXiv:2112.06861 [gr-qc]} \BibitemShut {NoStop}%
\bibitem [{\citenamefont {Aasi}\ \emph {et~al.}(2015)\citenamefont {Aasi} \emph {et~al.}}]{LIGOScientific:2014pky}%
  \BibitemOpen
  \bibfield  {author} {\bibinfo {author} {\bibfnamefont {J.}~\bibnamefont {Aasi}} \emph {et~al.} (\bibinfo {collaboration} {LIGO Scientific}),\ }\bibfield  {title} {\enquote {\bibinfo {title} {{Advanced LIGO}},}\ }\href {\doibase 10.1088/0264-9381/32/7/074001} {\bibfield  {journal} {\bibinfo  {journal} {Class. Quant. Grav.}\ }\textbf {\bibinfo {volume} {32}},\ \bibinfo {pages} {074001} (\bibinfo {year} {2015})},\ \Eprint {http://arxiv.org/abs/1411.4547} {arXiv:1411.4547 [gr-qc]} \BibitemShut {NoStop}%
\bibitem [{\citenamefont {Acernese}\ \emph {et~al.}(2015)\citenamefont {Acernese} \emph {et~al.}}]{VIRGO:2014yos}%
  \BibitemOpen
  \bibfield  {author} {\bibinfo {author} {\bibfnamefont {F.}~\bibnamefont {Acernese}} \emph {et~al.} (\bibinfo {collaboration} {VIRGO}),\ }\bibfield  {title} {\enquote {\bibinfo {title} {{Advanced Virgo: a second-generation interferometric gravitational wave detector}},}\ }\href {\doibase 10.1088/0264-9381/32/2/024001} {\bibfield  {journal} {\bibinfo  {journal} {Class. Quant. Grav.}\ }\textbf {\bibinfo {volume} {32}},\ \bibinfo {pages} {024001} (\bibinfo {year} {2015})},\ \Eprint {http://arxiv.org/abs/1408.3978} {arXiv:1408.3978 [gr-qc]} \BibitemShut {NoStop}%
\bibitem [{\citenamefont {Abbott}\ \emph {et~al.}(2019{\natexlab{b}})\citenamefont {Abbott} \emph {et~al.}}]{LIGOScientific:2018jsj}%
  \BibitemOpen
  \bibfield  {author} {\bibinfo {author} {\bibfnamefont {B.~P.}\ \bibnamefont {Abbott}} \emph {et~al.} (\bibinfo {collaboration} {LIGO Scientific, Virgo}),\ }\bibfield  {title} {\enquote {\bibinfo {title} {{Binary Black Hole Population Properties Inferred from the First and Second Observing Runs of Advanced LIGO and Advanced Virgo}},}\ }\href {\doibase 10.3847/2041-8213/ab3800} {\bibfield  {journal} {\bibinfo  {journal} {Astrophys. J. Lett.}\ }\textbf {\bibinfo {volume} {882}},\ \bibinfo {pages} {L24} (\bibinfo {year} {2019}{\natexlab{b}})},\ \Eprint {http://arxiv.org/abs/1811.12940} {arXiv:1811.12940 [astro-ph.HE]} \BibitemShut {NoStop}%
\bibitem [{\citenamefont {Abbott}\ \emph {et~al.}(2021{\natexlab{c}})\citenamefont {Abbott} \emph {et~al.}}]{LIGOScientific:2020kqk}%
  \BibitemOpen
  \bibfield  {author} {\bibinfo {author} {\bibfnamefont {R.}~\bibnamefont {Abbott}} \emph {et~al.} (\bibinfo {collaboration} {LIGO Scientific, Virgo}),\ }\bibfield  {title} {\enquote {\bibinfo {title} {{Population Properties of Compact Objects from the Second LIGO-Virgo Gravitational-Wave Transient Catalog}},}\ }\href {\doibase 10.3847/2041-8213/abe949} {\bibfield  {journal} {\bibinfo  {journal} {Astrophys. J. Lett.}\ }\textbf {\bibinfo {volume} {913}},\ \bibinfo {pages} {L7} (\bibinfo {year} {2021}{\natexlab{c}})},\ \Eprint {http://arxiv.org/abs/2010.14533} {arXiv:2010.14533 [astro-ph.HE]} \BibitemShut {NoStop}%
\bibitem [{\citenamefont {Abbott}\ \emph {et~al.}(2023)\citenamefont {Abbott} \emph {et~al.}}]{KAGRA:2021duu}%
  \BibitemOpen
  \bibfield  {author} {\bibinfo {author} {\bibfnamefont {R.}~\bibnamefont {Abbott}} \emph {et~al.} (\bibinfo {collaboration} {KAGRA, VIRGO, LIGO Scientific}),\ }\bibfield  {title} {\enquote {\bibinfo {title} {{Population of Merging Compact Binaries Inferred Using Gravitational Waves through GWTC-3}},}\ }\href {\doibase 10.1103/PhysRevX.13.011048} {\bibfield  {journal} {\bibinfo  {journal} {Phys. Rev. X}\ }\textbf {\bibinfo {volume} {13}},\ \bibinfo {pages} {011048} (\bibinfo {year} {2023})},\ \Eprint {http://arxiv.org/abs/2111.03634} {arXiv:2111.03634 [astro-ph.HE]} \BibitemShut {NoStop}%
\bibitem [{\citenamefont {Agazie}\ \emph {et~al.}(2023{\natexlab{a}})\citenamefont {Agazie} \emph {et~al.}}]{NANOGrav:2023hde}%
  \BibitemOpen
  \bibfield  {author} {\bibinfo {author} {\bibfnamefont {Gabriella}\ \bibnamefont {Agazie}} \emph {et~al.} (\bibinfo {collaboration} {NANOGrav}),\ }\bibfield  {title} {\enquote {\bibinfo {title} {{The NANOGrav 15 yr Data Set: Observations and Timing of 68 Millisecond Pulsars}},}\ }\href {\doibase 10.3847/2041-8213/acda9a} {\bibfield  {journal} {\bibinfo  {journal} {Astrophys. J. Lett.}\ }\textbf {\bibinfo {volume} {951}},\ \bibinfo {pages} {L9} (\bibinfo {year} {2023}{\natexlab{a}})},\ \Eprint {http://arxiv.org/abs/2306.16217} {arXiv:2306.16217 [astro-ph.HE]} \BibitemShut {NoStop}%
\bibitem [{\citenamefont {Agazie}\ \emph {et~al.}(2023{\natexlab{b}})\citenamefont {Agazie} \emph {et~al.}}]{NANOGrav:2023gor}%
  \BibitemOpen
  \bibfield  {author} {\bibinfo {author} {\bibfnamefont {Gabriella}\ \bibnamefont {Agazie}} \emph {et~al.} (\bibinfo {collaboration} {NANOGrav}),\ }\bibfield  {title} {\enquote {\bibinfo {title} {{The NANOGrav 15 yr Data Set: Evidence for a Gravitational-wave Background}},}\ }\href {\doibase 10.3847/2041-8213/acdac6} {\bibfield  {journal} {\bibinfo  {journal} {Astrophys. J. Lett.}\ }\textbf {\bibinfo {volume} {951}},\ \bibinfo {pages} {L8} (\bibinfo {year} {2023}{\natexlab{b}})},\ \Eprint {http://arxiv.org/abs/2306.16213} {arXiv:2306.16213 [astro-ph.HE]} \BibitemShut {NoStop}%
\bibitem [{\citenamefont {Zic}\ \emph {et~al.}(2023)\citenamefont {Zic} \emph {et~al.}}]{Zic:2023gta}%
  \BibitemOpen
  \bibfield  {author} {\bibinfo {author} {\bibfnamefont {Andrew}\ \bibnamefont {Zic}} \emph {et~al.},\ }\bibfield  {title} {\enquote {\bibinfo {title} {{The Parkes Pulsar Timing Array third data release}},}\ }\href {\doibase 10.1017/pasa.2023.36} {\bibfield  {journal} {\bibinfo  {journal} {Publ. Astron. Soc. Austral.}\ }\textbf {\bibinfo {volume} {40}},\ \bibinfo {pages} {e049} (\bibinfo {year} {2023})},\ \Eprint {http://arxiv.org/abs/2306.16230} {arXiv:2306.16230 [astro-ph.HE]} \BibitemShut {NoStop}%
\bibitem [{\citenamefont {Reardon}\ \emph {et~al.}(2023)\citenamefont {Reardon} \emph {et~al.}}]{Reardon:2023gzh}%
  \BibitemOpen
  \bibfield  {author} {\bibinfo {author} {\bibfnamefont {Daniel~J.}\ \bibnamefont {Reardon}} \emph {et~al.},\ }\bibfield  {title} {\enquote {\bibinfo {title} {{Search for an Isotropic Gravitational-wave Background with the Parkes Pulsar Timing Array}},}\ }\href {\doibase 10.3847/2041-8213/acdd02} {\bibfield  {journal} {\bibinfo  {journal} {Astrophys. J. Lett.}\ }\textbf {\bibinfo {volume} {951}},\ \bibinfo {pages} {L6} (\bibinfo {year} {2023})},\ \Eprint {http://arxiv.org/abs/2306.16215} {arXiv:2306.16215 [astro-ph.HE]} \BibitemShut {NoStop}%
\bibitem [{\citenamefont {Antoniadis}\ \emph {et~al.}(2023{\natexlab{a}})\citenamefont {Antoniadis} \emph {et~al.}}]{Antoniadis:2023lym}%
  \BibitemOpen
  \bibfield  {author} {\bibinfo {author} {\bibfnamefont {J.}~\bibnamefont {Antoniadis}} \emph {et~al.} (\bibinfo {collaboration} {EPTA}),\ }\bibfield  {title} {\enquote {\bibinfo {title} {{The second data release from the European Pulsar Timing Array - I. The dataset and timing analysis}},}\ }\href {\doibase 10.1051/0004-6361/202346841} {\bibfield  {journal} {\bibinfo  {journal} {Astron. Astrophys.}\ }\textbf {\bibinfo {volume} {678}},\ \bibinfo {pages} {A48} (\bibinfo {year} {2023}{\natexlab{a}})},\ \Eprint {http://arxiv.org/abs/2306.16224} {arXiv:2306.16224 [astro-ph.HE]} \BibitemShut {NoStop}%
\bibitem [{\citenamefont {Antoniadis}\ \emph {et~al.}(2023{\natexlab{b}})\citenamefont {Antoniadis} \emph {et~al.}}]{Antoniadis:2023ott}%
  \BibitemOpen
  \bibfield  {author} {\bibinfo {author} {\bibfnamefont {J.}~\bibnamefont {Antoniadis}} \emph {et~al.} (\bibinfo {collaboration} {EPTA, InPTA:}),\ }\bibfield  {title} {\enquote {\bibinfo {title} {{The second data release from the European Pulsar Timing Array - III. Search for gravitational wave signals}},}\ }\href {\doibase 10.1051/0004-6361/202346844} {\bibfield  {journal} {\bibinfo  {journal} {Astron. Astrophys.}\ }\textbf {\bibinfo {volume} {678}},\ \bibinfo {pages} {A50} (\bibinfo {year} {2023}{\natexlab{b}})},\ \Eprint {http://arxiv.org/abs/2306.16214} {arXiv:2306.16214 [astro-ph.HE]} \BibitemShut {NoStop}%
\bibitem [{\citenamefont {Xu}\ \emph {et~al.}(2023)\citenamefont {Xu} \emph {et~al.}}]{Xu:2023wog}%
  \BibitemOpen
  \bibfield  {author} {\bibinfo {author} {\bibfnamefont {Heng}\ \bibnamefont {Xu}} \emph {et~al.},\ }\bibfield  {title} {\enquote {\bibinfo {title} {{Searching for the Nano-Hertz Stochastic Gravitational Wave Background with the Chinese Pulsar Timing Array Data Release I}},}\ }\href {\doibase 10.1088/1674-4527/acdfa5} {\bibfield  {journal} {\bibinfo  {journal} {Res. Astron. Astrophys.}\ }\textbf {\bibinfo {volume} {23}},\ \bibinfo {pages} {075024} (\bibinfo {year} {2023})},\ \Eprint {http://arxiv.org/abs/2306.16216} {arXiv:2306.16216 [astro-ph.HE]} \BibitemShut {NoStop}%
\bibitem [{\citenamefont {Hellings}\ and\ \citenamefont {Downs}(1983)}]{Hellings:1983fr}%
  \BibitemOpen
  \bibfield  {author} {\bibinfo {author} {\bibfnamefont {R.~w.}\ \bibnamefont {Hellings}}\ and\ \bibinfo {author} {\bibfnamefont {G.~s.}\ \bibnamefont {Downs}},\ }\bibfield  {title} {\enquote {\bibinfo {title} {{UPPER LIMITS ON THE ISOTROPIC GRAVITATIONAL RADIATION BACKGROUND FROM PULSAR TIMING ANALYSIS}},}\ }\href {\doibase 10.1086/183954} {\bibfield  {journal} {\bibinfo  {journal} {Astrophys. J. Lett.}\ }\textbf {\bibinfo {volume} {265}},\ \bibinfo {pages} {L39--L42} (\bibinfo {year} {1983})}\BibitemShut {NoStop}%
\bibitem [{\citenamefont {Afzal}\ \emph {et~al.}(2023)\citenamefont {Afzal} \emph {et~al.}}]{NANOGrav:2023hvm}%
  \BibitemOpen
  \bibfield  {author} {\bibinfo {author} {\bibfnamefont {Adeela}\ \bibnamefont {Afzal}} \emph {et~al.} (\bibinfo {collaboration} {NANOGrav}),\ }\bibfield  {title} {\enquote {\bibinfo {title} {{The NANOGrav 15 yr Data Set: Search for Signals from New Physics}},}\ }\href {\doibase 10.3847/2041-8213/acdc91} {\bibfield  {journal} {\bibinfo  {journal} {Astrophys. J. Lett.}\ }\textbf {\bibinfo {volume} {951}},\ \bibinfo {pages} {L11} (\bibinfo {year} {2023})},\ \bibinfo {note} {[Erratum: Astrophys.J.Lett. 971, L27 (2024), Erratum: Astrophys.J. 971, L27 (2024)]},\ \Eprint {http://arxiv.org/abs/2306.16219} {arXiv:2306.16219 [astro-ph.HE]} \BibitemShut {NoStop}%
\bibitem [{\citenamefont {Antoniadis}\ \emph {et~al.}(2024)\citenamefont {Antoniadis} \emph {et~al.}}]{Antoniadis:2023xlr}%
  \BibitemOpen
  \bibfield  {author} {\bibinfo {author} {\bibfnamefont {J.}~\bibnamefont {Antoniadis}} \emph {et~al.} (\bibinfo {collaboration} {EPTA, InPTA}),\ }\bibfield  {title} {\enquote {\bibinfo {title} {{The second data release from the European Pulsar Timing Array - IV. Implications for massive black holes, dark matter, and the early Universe}},}\ }\href {\doibase 10.1051/0004-6361/202347433} {\bibfield  {journal} {\bibinfo  {journal} {Astron. Astrophys.}\ }\textbf {\bibinfo {volume} {685}},\ \bibinfo {pages} {A94} (\bibinfo {year} {2024})},\ \Eprint {http://arxiv.org/abs/2306.16227} {arXiv:2306.16227 [astro-ph.CO]} \BibitemShut {NoStop}%
\bibitem [{\citenamefont {Agazie}\ \emph {et~al.}(2023{\natexlab{c}})\citenamefont {Agazie} \emph {et~al.}}]{NANOGrav:2023hfp}%
  \BibitemOpen
  \bibfield  {author} {\bibinfo {author} {\bibfnamefont {Gabriella}\ \bibnamefont {Agazie}} \emph {et~al.} (\bibinfo {collaboration} {NANOGrav}),\ }\bibfield  {title} {\enquote {\bibinfo {title} {{The NANOGrav 15 yr Data Set: Constraints on Supermassive Black Hole Binaries from the Gravitational-wave Background}},}\ }\href {\doibase 10.3847/2041-8213/ace18b} {\bibfield  {journal} {\bibinfo  {journal} {Astrophys. J. Lett.}\ }\textbf {\bibinfo {volume} {952}},\ \bibinfo {pages} {L37} (\bibinfo {year} {2023}{\natexlab{c}})},\ \Eprint {http://arxiv.org/abs/2306.16220} {arXiv:2306.16220 [astro-ph.HE]} \BibitemShut {NoStop}%
\bibitem [{\citenamefont {Ellis}\ \emph {et~al.}(2024)\citenamefont {Ellis}, \citenamefont {Fairbairn}, \citenamefont {H\"utsi}, \citenamefont {Raidal}, \citenamefont {Urrutia}, \citenamefont {Vaskonen},\ and\ \citenamefont {Veerm\"ae}}]{Ellis:2023dgf}%
  \BibitemOpen
  \bibfield  {author} {\bibinfo {author} {\bibfnamefont {John}\ \bibnamefont {Ellis}}, \bibinfo {author} {\bibfnamefont {Malcolm}\ \bibnamefont {Fairbairn}}, \bibinfo {author} {\bibfnamefont {Gert}\ \bibnamefont {H\"utsi}}, \bibinfo {author} {\bibfnamefont {Juhan}\ \bibnamefont {Raidal}}, \bibinfo {author} {\bibfnamefont {Juan}\ \bibnamefont {Urrutia}}, \bibinfo {author} {\bibfnamefont {Ville}\ \bibnamefont {Vaskonen}}, \ and\ \bibinfo {author} {\bibfnamefont {Hardi}\ \bibnamefont {Veerm\"ae}},\ }\bibfield  {title} {\enquote {\bibinfo {title} {{Gravitational waves from supermassive black hole binaries in light of the NANOGrav 15-year data}},}\ }\href {\doibase 10.1103/PhysRevD.109.L021302} {\bibfield  {journal} {\bibinfo  {journal} {Phys. Rev. D}\ }\textbf {\bibinfo {volume} {109}},\ \bibinfo {pages} {L021302} (\bibinfo {year} {2024})},\ \Eprint {http://arxiv.org/abs/2306.17021} {arXiv:2306.17021 [astro-ph.CO]} \BibitemShut {NoStop}%
\bibitem [{\citenamefont {Shen}\ \emph {et~al.}(2023)\citenamefont {Shen}, \citenamefont {Yuan}, \citenamefont {Wang},\ and\ \citenamefont {Wang}}]{Shen:2023pan}%
  \BibitemOpen
  \bibfield  {author} {\bibinfo {author} {\bibfnamefont {Zhao-Qiang}\ \bibnamefont {Shen}}, \bibinfo {author} {\bibfnamefont {Guan-Wen}\ \bibnamefont {Yuan}}, \bibinfo {author} {\bibfnamefont {Yi-Ying}\ \bibnamefont {Wang}}, \ and\ \bibinfo {author} {\bibfnamefont {Yuan-Zhu}\ \bibnamefont {Wang}},\ }\bibfield  {title} {\enquote {\bibinfo {title} {{Dark Matter Spike surrounding Supermassive Black Holes Binary and the nanohertz Stochastic Gravitational Wave Background}},}\ }\href@noop {} {\  (\bibinfo {year} {2023})},\ \Eprint {http://arxiv.org/abs/2306.17143} {arXiv:2306.17143 [astro-ph.HE]} \BibitemShut {NoStop}%
\bibitem [{\citenamefont {Bi}\ \emph {et~al.}(2023)\citenamefont {Bi}, \citenamefont {Wu}, \citenamefont {Chen},\ and\ \citenamefont {Huang}}]{Bi:2023tib}%
  \BibitemOpen
  \bibfield  {author} {\bibinfo {author} {\bibfnamefont {Yan-Chen}\ \bibnamefont {Bi}}, \bibinfo {author} {\bibfnamefont {Yu-Mei}\ \bibnamefont {Wu}}, \bibinfo {author} {\bibfnamefont {Zu-Cheng}\ \bibnamefont {Chen}}, \ and\ \bibinfo {author} {\bibfnamefont {Qing-Guo}\ \bibnamefont {Huang}},\ }\bibfield  {title} {\enquote {\bibinfo {title} {{Implications for the supermassive black hole binaries from the NANOGrav 15-year data set}},}\ }\href {\doibase 10.1007/s11433-023-2252-4} {\bibfield  {journal} {\bibinfo  {journal} {Sci. China Phys. Mech. Astron.}\ }\textbf {\bibinfo {volume} {66}},\ \bibinfo {pages} {120402} (\bibinfo {year} {2023})},\ \Eprint {http://arxiv.org/abs/2307.00722} {arXiv:2307.00722 [astro-ph.CO]} \BibitemShut {NoStop}%
\bibitem [{\citenamefont {Barausse}\ \emph {et~al.}(2023)\citenamefont {Barausse}, \citenamefont {Dey}, \citenamefont {Crisostomi}, \citenamefont {Panayada}, \citenamefont {Marsat},\ and\ \citenamefont {Basak}}]{Barausse:2023yrx}%
  \BibitemOpen
  \bibfield  {author} {\bibinfo {author} {\bibfnamefont {Enrico}\ \bibnamefont {Barausse}}, \bibinfo {author} {\bibfnamefont {Kallol}\ \bibnamefont {Dey}}, \bibinfo {author} {\bibfnamefont {Marco}\ \bibnamefont {Crisostomi}}, \bibinfo {author} {\bibfnamefont {Akshay}\ \bibnamefont {Panayada}}, \bibinfo {author} {\bibfnamefont {Sylvain}\ \bibnamefont {Marsat}}, \ and\ \bibinfo {author} {\bibfnamefont {Soumen}\ \bibnamefont {Basak}},\ }\bibfield  {title} {\enquote {\bibinfo {title} {{Implications of the pulsar timing array detections for massive black hole mergers in the LISA band}},}\ }\href {\doibase 10.1103/PhysRevD.108.103034} {\bibfield  {journal} {\bibinfo  {journal} {Phys. Rev. D}\ }\textbf {\bibinfo {volume} {108}},\ \bibinfo {pages} {103034} (\bibinfo {year} {2023})},\ \Eprint {http://arxiv.org/abs/2307.12245} {arXiv:2307.12245 [astro-ph.GA]} \BibitemShut {NoStop}%
\bibitem [{\citenamefont {Liu}\ \emph {et~al.}(2024{\natexlab{a}})\citenamefont {Liu}, \citenamefont {Chen},\ and\ \citenamefont {Huang}}]{Liu:2023ymk}%
  \BibitemOpen
  \bibfield  {author} {\bibinfo {author} {\bibfnamefont {Lang}\ \bibnamefont {Liu}}, \bibinfo {author} {\bibfnamefont {Zu-Cheng}\ \bibnamefont {Chen}}, \ and\ \bibinfo {author} {\bibfnamefont {Qing-Guo}\ \bibnamefont {Huang}},\ }\bibfield  {title} {\enquote {\bibinfo {title} {{Implications for the non-Gaussianity of curvature perturbation from pulsar timing arrays}},}\ }\href {\doibase 10.1103/PhysRevD.109.L061301} {\bibfield  {journal} {\bibinfo  {journal} {Phys. Rev. D}\ }\textbf {\bibinfo {volume} {109}},\ \bibinfo {pages} {L061301} (\bibinfo {year} {2024}{\natexlab{a}})},\ \Eprint {http://arxiv.org/abs/2307.01102} {arXiv:2307.01102 [astro-ph.CO]} \BibitemShut {NoStop}%
\bibitem [{\citenamefont {Franciolini}\ \emph {et~al.}(2023)\citenamefont {Franciolini}, \citenamefont {Iovino}, \citenamefont {Vaskonen},\ and\ \citenamefont {Veermae}}]{Franciolini:2023pbf}%
  \BibitemOpen
  \bibfield  {author} {\bibinfo {author} {\bibfnamefont {Gabriele}\ \bibnamefont {Franciolini}}, \bibinfo {author} {\bibfnamefont {Antonio}\ \bibnamefont {Iovino}, \bibfnamefont {Junior.}}, \bibinfo {author} {\bibfnamefont {Ville}\ \bibnamefont {Vaskonen}}, \ and\ \bibinfo {author} {\bibfnamefont {Hardi}\ \bibnamefont {Veermae}},\ }\bibfield  {title} {\enquote {\bibinfo {title} {{Recent Gravitational Wave Observation by Pulsar Timing Arrays and Primordial Black Holes: The Importance of Non-Gaussianities}},}\ }\href {\doibase 10.1103/PhysRevLett.131.201401} {\bibfield  {journal} {\bibinfo  {journal} {Phys. Rev. Lett.}\ }\textbf {\bibinfo {volume} {131}},\ \bibinfo {pages} {201401} (\bibinfo {year} {2023})},\ \Eprint {http://arxiv.org/abs/2306.17149} {arXiv:2306.17149 [astro-ph.CO]} \BibitemShut {NoStop}%
\bibitem [{\citenamefont {Wang}\ \emph {et~al.}(2024{\natexlab{a}})\citenamefont {Wang}, \citenamefont {Zhao}, \citenamefont {Li},\ and\ \citenamefont {Zhu}}]{Wang:2023ost}%
  \BibitemOpen
  \bibfield  {author} {\bibinfo {author} {\bibfnamefont {Sai}\ \bibnamefont {Wang}}, \bibinfo {author} {\bibfnamefont {Zhi-Chao}\ \bibnamefont {Zhao}}, \bibinfo {author} {\bibfnamefont {Jun-Peng}\ \bibnamefont {Li}}, \ and\ \bibinfo {author} {\bibfnamefont {Qing-Hua}\ \bibnamefont {Zhu}},\ }\bibfield  {title} {\enquote {\bibinfo {title} {{Implications of pulsar timing array data for scalar-induced gravitational waves and primordial black holes: Primordial non-Gaussianity fNL considered}},}\ }\href {\doibase 10.1103/PhysRevResearch.6.L012060} {\bibfield  {journal} {\bibinfo  {journal} {Phys. Rev. Res.}\ }\textbf {\bibinfo {volume} {6}},\ \bibinfo {pages} {L012060} (\bibinfo {year} {2024}{\natexlab{a}})},\ \Eprint {http://arxiv.org/abs/2307.00572} {arXiv:2307.00572 [astro-ph.CO]} \BibitemShut {NoStop}%
\bibitem [{\citenamefont {Jin}\ \emph {et~al.}(2023)\citenamefont {Jin}, \citenamefont {Chen}, \citenamefont {Yi}, \citenamefont {You}, \citenamefont {Liu},\ and\ \citenamefont {Wu}}]{Jin:2023wri}%
  \BibitemOpen
  \bibfield  {author} {\bibinfo {author} {\bibfnamefont {Jia-Heng}\ \bibnamefont {Jin}}, \bibinfo {author} {\bibfnamefont {Zu-Cheng}\ \bibnamefont {Chen}}, \bibinfo {author} {\bibfnamefont {Zhu}\ \bibnamefont {Yi}}, \bibinfo {author} {\bibfnamefont {Zhi-Qiang}\ \bibnamefont {You}}, \bibinfo {author} {\bibfnamefont {Lang}\ \bibnamefont {Liu}}, \ and\ \bibinfo {author} {\bibfnamefont {You}\ \bibnamefont {Wu}},\ }\bibfield  {title} {\enquote {\bibinfo {title} {{Confronting sound speed resonance with pulsar timing arrays}},}\ }\href {\doibase 10.1088/1475-7516/2023/09/016} {\bibfield  {journal} {\bibinfo  {journal} {JCAP}\ }\textbf {\bibinfo {volume} {09}},\ \bibinfo {pages} {016} (\bibinfo {year} {2023})},\ \Eprint {http://arxiv.org/abs/2307.08687} {arXiv:2307.08687 [astro-ph.CO]} \BibitemShut {NoStop}%
\bibitem [{\citenamefont {Liu}\ \emph {et~al.}(2023{\natexlab{a}})\citenamefont {Liu}, \citenamefont {Chen},\ and\ \citenamefont {Huang}}]{Liu:2023pau}%
  \BibitemOpen
  \bibfield  {author} {\bibinfo {author} {\bibfnamefont {Lang}\ \bibnamefont {Liu}}, \bibinfo {author} {\bibfnamefont {Zu-Cheng}\ \bibnamefont {Chen}}, \ and\ \bibinfo {author} {\bibfnamefont {Qing-Guo}\ \bibnamefont {Huang}},\ }\bibfield  {title} {\enquote {\bibinfo {title} {{Probing the equation of state of the early Universe with pulsar timing arrays}},}\ }\href {\doibase 10.1088/1475-7516/2023/11/071} {\bibfield  {journal} {\bibinfo  {journal} {JCAP}\ }\textbf {\bibinfo {volume} {11}},\ \bibinfo {pages} {071} (\bibinfo {year} {2023}{\natexlab{a}})},\ \Eprint {http://arxiv.org/abs/2307.14911} {arXiv:2307.14911 [astro-ph.CO]} \BibitemShut {NoStop}%
\bibitem [{\citenamefont {Zhu}\ \emph {et~al.}(2023)\citenamefont {Zhu}, \citenamefont {Zhao}, \citenamefont {Wang},\ and\ \citenamefont {Zhang}}]{Zhao:2023joc}%
  \BibitemOpen
  \bibfield  {author} {\bibinfo {author} {\bibfnamefont {Qing-Hua}\ \bibnamefont {Zhu}}, \bibinfo {author} {\bibfnamefont {Zhi-Chao}\ \bibnamefont {Zhao}}, \bibinfo {author} {\bibfnamefont {Sai}\ \bibnamefont {Wang}}, \ and\ \bibinfo {author} {\bibfnamefont {Xin}\ \bibnamefont {Zhang}},\ }\bibfield  {title} {\enquote {\bibinfo {title} {{Unraveling the early universe's equation of state and primordial black hole production with PTA, BBN, and CMB observations}},}\ }\href@noop {} {\  (\bibinfo {year} {2023})},\ \Eprint {http://arxiv.org/abs/2307.13574} {arXiv:2307.13574 [astro-ph.CO]} \BibitemShut {NoStop}%
\bibitem [{\citenamefont {Yi}\ \emph {et~al.}(2024{\natexlab{a}})\citenamefont {Yi}, \citenamefont {You}, \citenamefont {Wu}, \citenamefont {Chen},\ and\ \citenamefont {Liu}}]{Yi:2023npi}%
  \BibitemOpen
  \bibfield  {author} {\bibinfo {author} {\bibfnamefont {Zhu}\ \bibnamefont {Yi}}, \bibinfo {author} {\bibfnamefont {Zhi-Qiang}\ \bibnamefont {You}}, \bibinfo {author} {\bibfnamefont {You}\ \bibnamefont {Wu}}, \bibinfo {author} {\bibfnamefont {Zu-Cheng}\ \bibnamefont {Chen}}, \ and\ \bibinfo {author} {\bibfnamefont {Lang}\ \bibnamefont {Liu}},\ }\bibfield  {title} {\enquote {\bibinfo {title} {{Exploring the NANOGrav signal and planet-mass primordial black holes through Higgs inflation}},}\ }\href {\doibase 10.1088/1475-7516/2024/06/043} {\bibfield  {journal} {\bibinfo  {journal} {JCAP}\ }\textbf {\bibinfo {volume} {06}},\ \bibinfo {pages} {043} (\bibinfo {year} {2024}{\natexlab{a}})},\ \Eprint {http://arxiv.org/abs/2308.14688} {arXiv:2308.14688 [astro-ph.CO]} \BibitemShut {NoStop}%
\bibitem [{\citenamefont {Harigaya}\ \emph {et~al.}(2023)\citenamefont {Harigaya}, \citenamefont {Inomata},\ and\ \citenamefont {Terada}}]{Harigaya:2023pmw}%
  \BibitemOpen
  \bibfield  {author} {\bibinfo {author} {\bibfnamefont {Keisuke}\ \bibnamefont {Harigaya}}, \bibinfo {author} {\bibfnamefont {Keisuke}\ \bibnamefont {Inomata}}, \ and\ \bibinfo {author} {\bibfnamefont {Takahiro}\ \bibnamefont {Terada}},\ }\bibfield  {title} {\enquote {\bibinfo {title} {{Induced gravitational waves with kination era for recent pulsar timing array signals}},}\ }\href {\doibase 10.1103/PhysRevD.108.123538} {\bibfield  {journal} {\bibinfo  {journal} {Phys. Rev. D}\ }\textbf {\bibinfo {volume} {108}},\ \bibinfo {pages} {123538} (\bibinfo {year} {2023})},\ \Eprint {http://arxiv.org/abs/2309.00228} {arXiv:2309.00228 [astro-ph.CO]} \BibitemShut {NoStop}%
\bibitem [{\citenamefont {Balaji}\ \emph {et~al.}(2023)\citenamefont {Balaji}, \citenamefont {Dom\`enech},\ and\ \citenamefont {Franciolini}}]{Balaji:2023ehk}%
  \BibitemOpen
  \bibfield  {author} {\bibinfo {author} {\bibfnamefont {Shyam}\ \bibnamefont {Balaji}}, \bibinfo {author} {\bibfnamefont {Guillem}\ \bibnamefont {Dom\`enech}}, \ and\ \bibinfo {author} {\bibfnamefont {Gabriele}\ \bibnamefont {Franciolini}},\ }\bibfield  {title} {\enquote {\bibinfo {title} {{Scalar-induced gravitational wave interpretation of PTA data: the role of scalar fluctuation propagation speed}},}\ }\href {\doibase 10.1088/1475-7516/2023/10/041} {\bibfield  {journal} {\bibinfo  {journal} {JCAP}\ }\textbf {\bibinfo {volume} {10}},\ \bibinfo {pages} {041} (\bibinfo {year} {2023})},\ \Eprint {http://arxiv.org/abs/2307.08552} {arXiv:2307.08552 [gr-qc]} \BibitemShut {NoStop}%
\bibitem [{\citenamefont {Yi}\ \emph {et~al.}(2024{\natexlab{b}})\citenamefont {Yi}, \citenamefont {You},\ and\ \citenamefont {Wu}}]{Yi:2023tdk}%
  \BibitemOpen
  \bibfield  {author} {\bibinfo {author} {\bibfnamefont {Zhu}\ \bibnamefont {Yi}}, \bibinfo {author} {\bibfnamefont {Zhi-Qiang}\ \bibnamefont {You}}, \ and\ \bibinfo {author} {\bibfnamefont {You}\ \bibnamefont {Wu}},\ }\bibfield  {title} {\enquote {\bibinfo {title} {{Model-independent reconstruction of the primordial curvature power spectrum from PTA data}},}\ }\href {\doibase 10.1088/1475-7516/2024/01/066} {\bibfield  {journal} {\bibinfo  {journal} {JCAP}\ }\textbf {\bibinfo {volume} {01}},\ \bibinfo {pages} {066} (\bibinfo {year} {2024}{\natexlab{b}})},\ \Eprint {http://arxiv.org/abs/2308.05632} {arXiv:2308.05632 [astro-ph.CO]} \BibitemShut {NoStop}%
\bibitem [{\citenamefont {You}\ \emph {et~al.}(2023)\citenamefont {You}, \citenamefont {Yi},\ and\ \citenamefont {Wu}}]{You:2023rmn}%
  \BibitemOpen
  \bibfield  {author} {\bibinfo {author} {\bibfnamefont {Zhi-Qiang}\ \bibnamefont {You}}, \bibinfo {author} {\bibfnamefont {Zhu}\ \bibnamefont {Yi}}, \ and\ \bibinfo {author} {\bibfnamefont {You}\ \bibnamefont {Wu}},\ }\bibfield  {title} {\enquote {\bibinfo {title} {{Constraints on primordial curvature power spectrum with pulsar timing arrays}},}\ }\href {\doibase 10.1088/1475-7516/2023/11/065} {\bibfield  {journal} {\bibinfo  {journal} {JCAP}\ }\textbf {\bibinfo {volume} {11}},\ \bibinfo {pages} {065} (\bibinfo {year} {2023})},\ \Eprint {http://arxiv.org/abs/2307.04419} {arXiv:2307.04419 [gr-qc]} \BibitemShut {NoStop}%
\bibitem [{\citenamefont {Liu}\ \emph {et~al.}(2024{\natexlab{b}})\citenamefont {Liu}, \citenamefont {Wu},\ and\ \citenamefont {Chen}}]{Liu:2023hpw}%
  \BibitemOpen
  \bibfield  {author} {\bibinfo {author} {\bibfnamefont {Lang}\ \bibnamefont {Liu}}, \bibinfo {author} {\bibfnamefont {You}\ \bibnamefont {Wu}}, \ and\ \bibinfo {author} {\bibfnamefont {Zu-Cheng}\ \bibnamefont {Chen}},\ }\bibfield  {title} {\enquote {\bibinfo {title} {{Simultaneously probing the sound speed and equation of state of the early Universe with pulsar timing arrays}},}\ }\href {\doibase 10.1088/1475-7516/2024/04/011} {\bibfield  {journal} {\bibinfo  {journal} {JCAP}\ }\textbf {\bibinfo {volume} {04}},\ \bibinfo {pages} {011} (\bibinfo {year} {2024}{\natexlab{b}})},\ \Eprint {http://arxiv.org/abs/2310.16500} {arXiv:2310.16500 [astro-ph.CO]} \BibitemShut {NoStop}%
\bibitem [{\citenamefont {Dom\`enech}\ \emph {et~al.}(2024)\citenamefont {Dom\`enech}, \citenamefont {Pi}, \citenamefont {Wang},\ and\ \citenamefont {Wang}}]{Domenech:2024rks}%
  \BibitemOpen
  \bibfield  {author} {\bibinfo {author} {\bibfnamefont {Guillem}\ \bibnamefont {Dom\`enech}}, \bibinfo {author} {\bibfnamefont {Shi}\ \bibnamefont {Pi}}, \bibinfo {author} {\bibfnamefont {Ao}~\bibnamefont {Wang}}, \ and\ \bibinfo {author} {\bibfnamefont {Jianing}\ \bibnamefont {Wang}},\ }\bibfield  {title} {\enquote {\bibinfo {title} {{Induced Gravitational Wave interpretation of PTA data: a complete study for general equation of state}},}\ }\href@noop {} {\  (\bibinfo {year} {2024})},\ \Eprint {http://arxiv.org/abs/2402.18965} {arXiv:2402.18965 [astro-ph.CO]} \BibitemShut {NoStop}%
\bibitem [{\citenamefont {Chen}\ and\ \citenamefont {Liu}(2024{\natexlab{a}})}]{Chen:2024twp}%
  \BibitemOpen
  \bibfield  {author} {\bibinfo {author} {\bibfnamefont {Zu-Cheng}\ \bibnamefont {Chen}}\ and\ \bibinfo {author} {\bibfnamefont {Lang}\ \bibnamefont {Liu}},\ }\bibfield  {title} {\enquote {\bibinfo {title} {{Can we distinguish the adiabatic fluctuations and isocurvature fluctuations with pulsar timing arrays?}}}\ }\href@noop {} {\  (\bibinfo {year} {2024}{\natexlab{a}})},\ \Eprint {http://arxiv.org/abs/2402.16781} {arXiv:2402.16781 [astro-ph.CO]} \BibitemShut {NoStop}%
\bibitem [{\citenamefont {Chen}\ \emph {et~al.}(2024{\natexlab{a}})\citenamefont {Chen}, \citenamefont {Li}, \citenamefont {Liu},\ and\ \citenamefont {Yi}}]{Chen:2024fir}%
  \BibitemOpen
  \bibfield  {author} {\bibinfo {author} {\bibfnamefont {Zu-Cheng}\ \bibnamefont {Chen}}, \bibinfo {author} {\bibfnamefont {Jun}\ \bibnamefont {Li}}, \bibinfo {author} {\bibfnamefont {Lang}\ \bibnamefont {Liu}}, \ and\ \bibinfo {author} {\bibfnamefont {Zhu}\ \bibnamefont {Yi}},\ }\bibfield  {title} {\enquote {\bibinfo {title} {{Probing the speed of scalar-induced gravitational waves with pulsar timing arrays}},}\ }\href {\doibase 10.1103/PhysRevD.109.L101302} {\bibfield  {journal} {\bibinfo  {journal} {Phys. Rev. D}\ }\textbf {\bibinfo {volume} {109}},\ \bibinfo {pages} {L101302} (\bibinfo {year} {2024}{\natexlab{a}})},\ \Eprint {http://arxiv.org/abs/2401.09818} {arXiv:2401.09818 [gr-qc]} \BibitemShut {NoStop}%
\bibitem [{\citenamefont {Bhaumik}\ \emph {et~al.}(2023)\citenamefont {Bhaumik}, \citenamefont {Jain},\ and\ \citenamefont {Lewicki}}]{Bhaumik:2023wmw}%
  \BibitemOpen
  \bibfield  {author} {\bibinfo {author} {\bibfnamefont {Nilanjandev}\ \bibnamefont {Bhaumik}}, \bibinfo {author} {\bibfnamefont {Rajeev~Kumar}\ \bibnamefont {Jain}}, \ and\ \bibinfo {author} {\bibfnamefont {Marek}\ \bibnamefont {Lewicki}},\ }\bibfield  {title} {\enquote {\bibinfo {title} {{Ultralow mass primordial black holes in the early Universe can explain the pulsar timing array signal}},}\ }\href {\doibase 10.1103/PhysRevD.108.123532} {\bibfield  {journal} {\bibinfo  {journal} {Phys. Rev. D}\ }\textbf {\bibinfo {volume} {108}},\ \bibinfo {pages} {123532} (\bibinfo {year} {2023})},\ \Eprint {http://arxiv.org/abs/2308.07912} {arXiv:2308.07912 [astro-ph.CO]} \BibitemShut {NoStop}%
\bibitem [{\citenamefont {Liu}\ \emph {et~al.}(2020)\citenamefont {Liu}, \citenamefont {Guo}, \citenamefont {Cai},\ and\ \citenamefont {Kim}}]{Liu:2020cds}%
  \BibitemOpen
  \bibfield  {author} {\bibinfo {author} {\bibfnamefont {Lang}\ \bibnamefont {Liu}}, \bibinfo {author} {\bibfnamefont {Zong-Kuan}\ \bibnamefont {Guo}}, \bibinfo {author} {\bibfnamefont {Rong-Gen}\ \bibnamefont {Cai}}, \ and\ \bibinfo {author} {\bibfnamefont {Sang~Pyo}\ \bibnamefont {Kim}},\ }\bibfield  {title} {\enquote {\bibinfo {title} {{Merger rate distribution of primordial black hole binaries with electric charges}},}\ }\href {\doibase 10.1103/PhysRevD.102.043508} {\bibfield  {journal} {\bibinfo  {journal} {Phys. Rev. D}\ }\textbf {\bibinfo {volume} {102}},\ \bibinfo {pages} {043508} (\bibinfo {year} {2020})},\ \Eprint {http://arxiv.org/abs/2001.02984} {arXiv:2001.02984 [astro-ph.CO]} \BibitemShut {NoStop}%
\bibitem [{\citenamefont {Bousder}\ \emph {et~al.}(2023)\citenamefont {Bousder}, \citenamefont {Riadsolh}, \citenamefont {Fatimy}, \citenamefont {Belkacemi},\ and\ \citenamefont {Ez-Zahraouy}}]{Bousder:2023ida}%
  \BibitemOpen
  \bibfield  {author} {\bibinfo {author} {\bibfnamefont {M.}~\bibnamefont {Bousder}}, \bibinfo {author} {\bibfnamefont {A.}~\bibnamefont {Riadsolh}}, \bibinfo {author} {\bibfnamefont {A.~El}\ \bibnamefont {Fatimy}}, \bibinfo {author} {\bibfnamefont {M.~El}\ \bibnamefont {Belkacemi}}, \ and\ \bibinfo {author} {\bibfnamefont {H.}~\bibnamefont {Ez-Zahraouy}},\ }\bibfield  {title} {\enquote {\bibinfo {title} {{Implications of the NANOGrav results for primordial black holes and Hubble tension}},}\ }\href@noop {} {\  (\bibinfo {year} {2023})},\ \Eprint {http://arxiv.org/abs/2307.10940} {arXiv:2307.10940 [gr-qc]} \BibitemShut {NoStop}%
\bibitem [{\citenamefont {Gouttenoire}\ \emph {et~al.}(2024)\citenamefont {Gouttenoire}, \citenamefont {Trifinopoulos}, \citenamefont {Valogiannis},\ and\ \citenamefont {Vanvlasselaer}}]{Gouttenoire:2023nzr}%
  \BibitemOpen
  \bibfield  {author} {\bibinfo {author} {\bibfnamefont {Yann}\ \bibnamefont {Gouttenoire}}, \bibinfo {author} {\bibfnamefont {Sokratis}\ \bibnamefont {Trifinopoulos}}, \bibinfo {author} {\bibfnamefont {Georgios}\ \bibnamefont {Valogiannis}}, \ and\ \bibinfo {author} {\bibfnamefont {Miguel}\ \bibnamefont {Vanvlasselaer}},\ }\bibfield  {title} {\enquote {\bibinfo {title} {{Scrutinizing the primordial black hole interpretation of PTA gravitational waves and JWST early galaxies}},}\ }\href {\doibase 10.1103/PhysRevD.109.123002} {\bibfield  {journal} {\bibinfo  {journal} {Phys. Rev. D}\ }\textbf {\bibinfo {volume} {109}},\ \bibinfo {pages} {123002} (\bibinfo {year} {2024})},\ \Eprint {http://arxiv.org/abs/2307.01457} {arXiv:2307.01457 [astro-ph.CO]} \BibitemShut {NoStop}%
\bibitem [{\citenamefont {Huang}\ \emph {et~al.}(2023)\citenamefont {Huang}, \citenamefont {Cai}, \citenamefont {Jiang}, \citenamefont {Zhang},\ and\ \citenamefont {Piao}}]{Huang:2023chx}%
  \BibitemOpen
  \bibfield  {author} {\bibinfo {author} {\bibfnamefont {Hai-Long}\ \bibnamefont {Huang}}, \bibinfo {author} {\bibfnamefont {Yong}\ \bibnamefont {Cai}}, \bibinfo {author} {\bibfnamefont {Jun-Qian}\ \bibnamefont {Jiang}}, \bibinfo {author} {\bibfnamefont {Jun}\ \bibnamefont {Zhang}}, \ and\ \bibinfo {author} {\bibfnamefont {Yun-Song}\ \bibnamefont {Piao}},\ }\bibfield  {title} {\enquote {\bibinfo {title} {{Supermassive primordial black holes in multiverse: for nano-Hertz gravitational wave and high-redshift JWST galaxies}},}\ }\href@noop {} {\  (\bibinfo {year} {2023})},\ \Eprint {http://arxiv.org/abs/2306.17577} {arXiv:2306.17577 [gr-qc]} \BibitemShut {NoStop}%
\bibitem [{\citenamefont {Depta}\ \emph {et~al.}(2023)\citenamefont {Depta}, \citenamefont {Schmidt-Hoberg}, \citenamefont {Schwaller},\ and\ \citenamefont {Tasillo}}]{Depta:2023qst}%
  \BibitemOpen
  \bibfield  {author} {\bibinfo {author} {\bibfnamefont {Paul~Frederik}\ \bibnamefont {Depta}}, \bibinfo {author} {\bibfnamefont {Kai}\ \bibnamefont {Schmidt-Hoberg}}, \bibinfo {author} {\bibfnamefont {Pedro}\ \bibnamefont {Schwaller}}, \ and\ \bibinfo {author} {\bibfnamefont {Carlo}\ \bibnamefont {Tasillo}},\ }\bibfield  {title} {\enquote {\bibinfo {title} {{Do pulsar timing arrays observe merging primordial black holes?}}}\ }\href@noop {} {\  (\bibinfo {year} {2023})},\ \Eprint {http://arxiv.org/abs/2306.17836} {arXiv:2306.17836 [astro-ph.CO]} \BibitemShut {NoStop}%
\bibitem [{\citenamefont {Wang}\ \emph {et~al.}(2024{\natexlab{b}})\citenamefont {Wang}, \citenamefont {Zhang},\ and\ \citenamefont {Sasaki}}]{Wang:2024vfv}%
  \BibitemOpen
  \bibfield  {author} {\bibinfo {author} {\bibfnamefont {Xinpeng}\ \bibnamefont {Wang}}, \bibinfo {author} {\bibfnamefont {Ying-li}\ \bibnamefont {Zhang}}, \ and\ \bibinfo {author} {\bibfnamefont {Misao}\ \bibnamefont {Sasaki}},\ }\bibfield  {title} {\enquote {\bibinfo {title} {{Enhanced curvature perturbation and primordial black hole formation in two-stage inflation with a break}},}\ }\href {\doibase 10.1088/1475-7516/2024/07/076} {\bibfield  {journal} {\bibinfo  {journal} {JCAP}\ }\textbf {\bibinfo {volume} {07}},\ \bibinfo {pages} {076} (\bibinfo {year} {2024}{\natexlab{b}})},\ \Eprint {http://arxiv.org/abs/2404.02492} {arXiv:2404.02492 [astro-ph.CO]} \BibitemShut {NoStop}%
\bibitem [{\citenamefont {Addazi}\ \emph {et~al.}(2024)\citenamefont {Addazi}, \citenamefont {Cai}, \citenamefont {Marciano},\ and\ \citenamefont {Visinelli}}]{Addazi:2023jvg}%
  \BibitemOpen
  \bibfield  {author} {\bibinfo {author} {\bibfnamefont {Andrea}\ \bibnamefont {Addazi}}, \bibinfo {author} {\bibfnamefont {Yi-Fu}\ \bibnamefont {Cai}}, \bibinfo {author} {\bibfnamefont {Antonino}\ \bibnamefont {Marciano}}, \ and\ \bibinfo {author} {\bibfnamefont {Luca}\ \bibnamefont {Visinelli}},\ }\bibfield  {title} {\enquote {\bibinfo {title} {{Have pulsar timing array methods detected a cosmological phase transition?}}}\ }\href {\doibase 10.1103/PhysRevD.109.015028} {\bibfield  {journal} {\bibinfo  {journal} {Phys. Rev. D}\ }\textbf {\bibinfo {volume} {109}},\ \bibinfo {pages} {015028} (\bibinfo {year} {2024})},\ \Eprint {http://arxiv.org/abs/2306.17205} {arXiv:2306.17205 [astro-ph.CO]} \BibitemShut {NoStop}%
\bibitem [{\citenamefont {Athron}\ \emph {et~al.}(2024)\citenamefont {Athron}, \citenamefont {Fowlie}, \citenamefont {Lu}, \citenamefont {Morris}, \citenamefont {Wu}, \citenamefont {Wu},\ and\ \citenamefont {Xu}}]{Athron:2023mer}%
  \BibitemOpen
  \bibfield  {author} {\bibinfo {author} {\bibfnamefont {Peter}\ \bibnamefont {Athron}}, \bibinfo {author} {\bibfnamefont {Andrew}\ \bibnamefont {Fowlie}}, \bibinfo {author} {\bibfnamefont {Chih-Ting}\ \bibnamefont {Lu}}, \bibinfo {author} {\bibfnamefont {Lachlan}\ \bibnamefont {Morris}}, \bibinfo {author} {\bibfnamefont {Lei}\ \bibnamefont {Wu}}, \bibinfo {author} {\bibfnamefont {Yongcheng}\ \bibnamefont {Wu}}, \ and\ \bibinfo {author} {\bibfnamefont {Zhongxiu}\ \bibnamefont {Xu}},\ }\bibfield  {title} {\enquote {\bibinfo {title} {{Can Supercooled Phase Transitions Explain the Gravitational Wave Background Observed by Pulsar Timing Arrays?}}}\ }\href {\doibase 10.1103/PhysRevLett.132.221001} {\bibfield  {journal} {\bibinfo  {journal} {Phys. Rev. Lett.}\ }\textbf {\bibinfo {volume} {132}},\ \bibinfo {pages} {221001} (\bibinfo {year} {2024})},\ \Eprint {http://arxiv.org/abs/2306.17239} {arXiv:2306.17239 [hep-ph]} \BibitemShut {NoStop}%
\bibitem [{\citenamefont {Zu}\ \emph {et~al.}(2024)\citenamefont {Zu}, \citenamefont {Zhang}, \citenamefont {Li}, \citenamefont {Gu}, \citenamefont {Tsai},\ and\ \citenamefont {Fan}}]{Zu:2023olm}%
  \BibitemOpen
  \bibfield  {author} {\bibinfo {author} {\bibfnamefont {Lei}\ \bibnamefont {Zu}}, \bibinfo {author} {\bibfnamefont {Chi}\ \bibnamefont {Zhang}}, \bibinfo {author} {\bibfnamefont {Yao-Yu}\ \bibnamefont {Li}}, \bibinfo {author} {\bibfnamefont {Yuchao}\ \bibnamefont {Gu}}, \bibinfo {author} {\bibfnamefont {Yue-Lin~Sming}\ \bibnamefont {Tsai}}, \ and\ \bibinfo {author} {\bibfnamefont {Yi-Zhong}\ \bibnamefont {Fan}},\ }\bibfield  {title} {\enquote {\bibinfo {title} {{Mirror QCD phase transition as the origin of the nanohertz Stochastic Gravitational-Wave Background}},}\ }\href {\doibase 10.1016/j.scib.2024.01.037} {\bibfield  {journal} {\bibinfo  {journal} {Sci. Bull.}\ }\textbf {\bibinfo {volume} {69}},\ \bibinfo {pages} {741--746} (\bibinfo {year} {2024})},\ \Eprint {http://arxiv.org/abs/2306.16769} {arXiv:2306.16769 [astro-ph.HE]} \BibitemShut {NoStop}%
\bibitem [{\citenamefont {Jiang}\ \emph {et~al.}(2024)\citenamefont {Jiang}, \citenamefont {Yang}, \citenamefont {Ma},\ and\ \citenamefont {Huang}}]{Jiang:2023qbm}%
  \BibitemOpen
  \bibfield  {author} {\bibinfo {author} {\bibfnamefont {Siyu}\ \bibnamefont {Jiang}}, \bibinfo {author} {\bibfnamefont {Aidi}\ \bibnamefont {Yang}}, \bibinfo {author} {\bibfnamefont {Jiucheng}\ \bibnamefont {Ma}}, \ and\ \bibinfo {author} {\bibfnamefont {Fa~Peng}\ \bibnamefont {Huang}},\ }\bibfield  {title} {\enquote {\bibinfo {title} {{Implication of nano-Hertz stochastic gravitational wave on dynamical dark matter through a dark first-order phase transition}},}\ }\href {\doibase 10.1088/1361-6382/ad24c6} {\bibfield  {journal} {\bibinfo  {journal} {Class. Quant. Grav.}\ }\textbf {\bibinfo {volume} {41}},\ \bibinfo {pages} {065009} (\bibinfo {year} {2024})},\ \Eprint {http://arxiv.org/abs/2306.17827} {arXiv:2306.17827 [hep-ph]} \BibitemShut {NoStop}%
\bibitem [{\citenamefont {Xiao}\ \emph {et~al.}(2023)\citenamefont {Xiao}, \citenamefont {Yang},\ and\ \citenamefont {Zhang}}]{Xiao:2023dbb}%
  \BibitemOpen
  \bibfield  {author} {\bibinfo {author} {\bibfnamefont {Yang}\ \bibnamefont {Xiao}}, \bibinfo {author} {\bibfnamefont {Jin~Min}\ \bibnamefont {Yang}}, \ and\ \bibinfo {author} {\bibfnamefont {Yang}\ \bibnamefont {Zhang}},\ }\bibfield  {title} {\enquote {\bibinfo {title} {{Implications of nano-Hertz gravitational waves on electroweak phase transition in the singlet dark matter model}},}\ }\href {\doibase 10.1016/j.scib.2023.11.025} {\bibfield  {journal} {\bibinfo  {journal} {Sci. Bull.}\ }\textbf {\bibinfo {volume} {68}},\ \bibinfo {pages} {3158--3164} (\bibinfo {year} {2023})},\ \Eprint {http://arxiv.org/abs/2307.01072} {arXiv:2307.01072 [hep-ph]} \BibitemShut {NoStop}%
\bibitem [{\citenamefont {Abe}\ and\ \citenamefont {Tada}(2023)}]{Abe:2023yrw}%
  \BibitemOpen
  \bibfield  {author} {\bibinfo {author} {\bibfnamefont {Katsuya~T.}\ \bibnamefont {Abe}}\ and\ \bibinfo {author} {\bibfnamefont {Yuichiro}\ \bibnamefont {Tada}},\ }\bibfield  {title} {\enquote {\bibinfo {title} {{Translating nano-Hertz gravitational wave background into primordial perturbations taking account of the cosmological QCD phase transition}},}\ }\href {\doibase 10.1103/PhysRevD.108.L101304} {\bibfield  {journal} {\bibinfo  {journal} {Phys. Rev. D}\ }\textbf {\bibinfo {volume} {108}},\ \bibinfo {pages} {L101304} (\bibinfo {year} {2023})},\ \Eprint {http://arxiv.org/abs/2307.01653} {arXiv:2307.01653 [astro-ph.CO]} \BibitemShut {NoStop}%
\bibitem [{\citenamefont {Gouttenoire}(2023)}]{Gouttenoire:2023bqy}%
  \BibitemOpen
  \bibfield  {author} {\bibinfo {author} {\bibfnamefont {Yann}\ \bibnamefont {Gouttenoire}},\ }\bibfield  {title} {\enquote {\bibinfo {title} {{First-Order Phase Transition Interpretation of Pulsar Timing Array Signal Is Consistent with Solar-Mass Black Holes}},}\ }\href {\doibase 10.1103/PhysRevLett.131.171404} {\bibfield  {journal} {\bibinfo  {journal} {Phys. Rev. Lett.}\ }\textbf {\bibinfo {volume} {131}},\ \bibinfo {pages} {171404} (\bibinfo {year} {2023})},\ \Eprint {http://arxiv.org/abs/2307.04239} {arXiv:2307.04239 [hep-ph]} \BibitemShut {NoStop}%
\bibitem [{\citenamefont {An}\ \emph {et~al.}(2024)\citenamefont {An}, \citenamefont {Su}, \citenamefont {Tai}, \citenamefont {Wang},\ and\ \citenamefont {Yang}}]{An:2023jxf}%
  \BibitemOpen
  \bibfield  {author} {\bibinfo {author} {\bibfnamefont {Haipeng}\ \bibnamefont {An}}, \bibinfo {author} {\bibfnamefont {Boye}\ \bibnamefont {Su}}, \bibinfo {author} {\bibfnamefont {Hanwen}\ \bibnamefont {Tai}}, \bibinfo {author} {\bibfnamefont {Lian-Tao}\ \bibnamefont {Wang}}, \ and\ \bibinfo {author} {\bibfnamefont {Chen}\ \bibnamefont {Yang}},\ }\bibfield  {title} {\enquote {\bibinfo {title} {{Phase transition during inflation and the gravitational wave signal at pulsar timing arrays}},}\ }\href {\doibase 10.1103/PhysRevD.109.L121304} {\bibfield  {journal} {\bibinfo  {journal} {Phys. Rev. D}\ }\textbf {\bibinfo {volume} {109}},\ \bibinfo {pages} {L121304} (\bibinfo {year} {2024})},\ \Eprint {http://arxiv.org/abs/2308.00070} {arXiv:2308.00070 [astro-ph.CO]} \BibitemShut {NoStop}%
\bibitem [{\citenamefont {Chen}\ \emph {et~al.}(2024{\natexlab{b}})\citenamefont {Chen}, \citenamefont {Li}, \citenamefont {Wu},\ and\ \citenamefont {Yu}}]{Chen:2023bms}%
  \BibitemOpen
  \bibfield  {author} {\bibinfo {author} {\bibfnamefont {Zu-Cheng}\ \bibnamefont {Chen}}, \bibinfo {author} {\bibfnamefont {Shou-Long}\ \bibnamefont {Li}}, \bibinfo {author} {\bibfnamefont {Puxun}\ \bibnamefont {Wu}}, \ and\ \bibinfo {author} {\bibfnamefont {Hongwei}\ \bibnamefont {Yu}},\ }\bibfield  {title} {\enquote {\bibinfo {title} {{NANOGrav hints for first-order confinement-deconfinement phase transition in different QCD-matter scenarios}},}\ }\href {\doibase 10.1103/PhysRevD.109.043022} {\bibfield  {journal} {\bibinfo  {journal} {Phys. Rev. D}\ }\textbf {\bibinfo {volume} {109}},\ \bibinfo {pages} {043022} (\bibinfo {year} {2024}{\natexlab{b}})},\ \Eprint {http://arxiv.org/abs/2312.01824} {arXiv:2312.01824 [astro-ph.CO]} \BibitemShut {NoStop}%
\bibitem [{\citenamefont {Kitajima}\ \emph {et~al.}(2024)\citenamefont {Kitajima}, \citenamefont {Lee}, \citenamefont {Murai}, \citenamefont {Takahashi},\ and\ \citenamefont {Yin}}]{Kitajima:2023cek}%
  \BibitemOpen
  \bibfield  {author} {\bibinfo {author} {\bibfnamefont {Naoya}\ \bibnamefont {Kitajima}}, \bibinfo {author} {\bibfnamefont {Junseok}\ \bibnamefont {Lee}}, \bibinfo {author} {\bibfnamefont {Kai}\ \bibnamefont {Murai}}, \bibinfo {author} {\bibfnamefont {Fuminobu}\ \bibnamefont {Takahashi}}, \ and\ \bibinfo {author} {\bibfnamefont {Wen}\ \bibnamefont {Yin}},\ }\bibfield  {title} {\enquote {\bibinfo {title} {{Gravitational waves from domain wall collapse, and application to nanohertz signals with QCD-coupled axions}},}\ }\href {\doibase 10.1016/j.physletb.2024.138586} {\bibfield  {journal} {\bibinfo  {journal} {Phys. Lett. B}\ }\textbf {\bibinfo {volume} {851}},\ \bibinfo {pages} {138586} (\bibinfo {year} {2024})},\ \Eprint {http://arxiv.org/abs/2306.17146} {arXiv:2306.17146 [hep-ph]} \BibitemShut {NoStop}%
\bibitem [{\citenamefont {Blasi}\ \emph {et~al.}(2023)\citenamefont {Blasi}, \citenamefont {Mariotti}, \citenamefont {Rase},\ and\ \citenamefont {Sevrin}}]{Blasi:2023sej}%
  \BibitemOpen
  \bibfield  {author} {\bibinfo {author} {\bibfnamefont {Simone}\ \bibnamefont {Blasi}}, \bibinfo {author} {\bibfnamefont {Alberto}\ \bibnamefont {Mariotti}}, \bibinfo {author} {\bibfnamefont {A\"aron}\ \bibnamefont {Rase}}, \ and\ \bibinfo {author} {\bibfnamefont {Alexander}\ \bibnamefont {Sevrin}},\ }\bibfield  {title} {\enquote {\bibinfo {title} {{Axionic domain walls at Pulsar Timing Arrays: QCD bias and particle friction}},}\ }\href {\doibase 10.1007/JHEP11(2023)169} {\bibfield  {journal} {\bibinfo  {journal} {JHEP}\ }\textbf {\bibinfo {volume} {11}},\ \bibinfo {pages} {169} (\bibinfo {year} {2023})},\ \Eprint {http://arxiv.org/abs/2306.17830} {arXiv:2306.17830 [hep-ph]} \BibitemShut {NoStop}%
\bibitem [{\citenamefont {Babichev}\ \emph {et~al.}(2023)\citenamefont {Babichev}, \citenamefont {Gorbunov}, \citenamefont {Ramazanov}, \citenamefont {Samanta},\ and\ \citenamefont {Vikman}}]{Babichev:2023pbf}%
  \BibitemOpen
  \bibfield  {author} {\bibinfo {author} {\bibfnamefont {E.}~\bibnamefont {Babichev}}, \bibinfo {author} {\bibfnamefont {D.}~\bibnamefont {Gorbunov}}, \bibinfo {author} {\bibfnamefont {S.}~\bibnamefont {Ramazanov}}, \bibinfo {author} {\bibfnamefont {R.}~\bibnamefont {Samanta}}, \ and\ \bibinfo {author} {\bibfnamefont {A.}~\bibnamefont {Vikman}},\ }\bibfield  {title} {\enquote {\bibinfo {title} {{NANOGrav spectral index \ensuremath{\gamma}=3 from melting domain walls}},}\ }\href {\doibase 10.1103/PhysRevD.108.123529} {\bibfield  {journal} {\bibinfo  {journal} {Phys. Rev. D}\ }\textbf {\bibinfo {volume} {108}},\ \bibinfo {pages} {123529} (\bibinfo {year} {2023})},\ \Eprint {http://arxiv.org/abs/2307.04582} {arXiv:2307.04582 [hep-ph]} \BibitemShut {NoStop}%
\bibitem [{\citenamefont {Kitajima}\ and\ \citenamefont {Nakayama}(2023)}]{Kitajima:2023vre}%
  \BibitemOpen
  \bibfield  {author} {\bibinfo {author} {\bibfnamefont {Naoya}\ \bibnamefont {Kitajima}}\ and\ \bibinfo {author} {\bibfnamefont {Kazunori}\ \bibnamefont {Nakayama}},\ }\bibfield  {title} {\enquote {\bibinfo {title} {{Nanohertz gravitational waves from cosmic strings and dark photon dark matter}},}\ }\href {\doibase 10.1016/j.physletb.2023.138213} {\bibfield  {journal} {\bibinfo  {journal} {Phys. Lett. B}\ }\textbf {\bibinfo {volume} {846}},\ \bibinfo {pages} {138213} (\bibinfo {year} {2023})},\ \Eprint {http://arxiv.org/abs/2306.17390} {arXiv:2306.17390 [hep-ph]} \BibitemShut {NoStop}%
\bibitem [{\citenamefont {Ellis}\ \emph {et~al.}(2023)\citenamefont {Ellis}, \citenamefont {Lewicki}, \citenamefont {Lin},\ and\ \citenamefont {Vaskonen}}]{Ellis:2023tsl}%
  \BibitemOpen
  \bibfield  {author} {\bibinfo {author} {\bibfnamefont {John}\ \bibnamefont {Ellis}}, \bibinfo {author} {\bibfnamefont {Marek}\ \bibnamefont {Lewicki}}, \bibinfo {author} {\bibfnamefont {Chunshan}\ \bibnamefont {Lin}}, \ and\ \bibinfo {author} {\bibfnamefont {Ville}\ \bibnamefont {Vaskonen}},\ }\bibfield  {title} {\enquote {\bibinfo {title} {{Cosmic superstrings revisited in light of NANOGrav 15-year data}},}\ }\href {\doibase 10.1103/PhysRevD.108.103511} {\bibfield  {journal} {\bibinfo  {journal} {Phys. Rev. D}\ }\textbf {\bibinfo {volume} {108}},\ \bibinfo {pages} {103511} (\bibinfo {year} {2023})},\ \Eprint {http://arxiv.org/abs/2306.17147} {arXiv:2306.17147 [astro-ph.CO]} \BibitemShut {NoStop}%
\bibitem [{\citenamefont {Wang}\ \emph {et~al.}(2023)\citenamefont {Wang}, \citenamefont {Lei}, \citenamefont {Jiao}, \citenamefont {Feng},\ and\ \citenamefont {Fan}}]{Wang:2023len}%
  \BibitemOpen
  \bibfield  {author} {\bibinfo {author} {\bibfnamefont {Ziwei}\ \bibnamefont {Wang}}, \bibinfo {author} {\bibfnamefont {Lei}\ \bibnamefont {Lei}}, \bibinfo {author} {\bibfnamefont {Hao}\ \bibnamefont {Jiao}}, \bibinfo {author} {\bibfnamefont {Lei}\ \bibnamefont {Feng}}, \ and\ \bibinfo {author} {\bibfnamefont {Yi-Zhong}\ \bibnamefont {Fan}},\ }\bibfield  {title} {\enquote {\bibinfo {title} {{The nanohertz stochastic gravitational wave background from cosmic string loops and the abundant high redshift massive galaxies}},}\ }\href {\doibase 10.1007/s11433-023-2262-0} {\bibfield  {journal} {\bibinfo  {journal} {Sci. China Phys. Mech. Astron.}\ }\textbf {\bibinfo {volume} {66}},\ \bibinfo {pages} {120403} (\bibinfo {year} {2023})},\ \Eprint {http://arxiv.org/abs/2306.17150} {arXiv:2306.17150 [astro-ph.HE]} \BibitemShut {NoStop}%
\bibitem [{\citenamefont {Ahmed}\ \emph {et~al.}(2024)\citenamefont {Ahmed}, \citenamefont {Chowdhury}, \citenamefont {Nasri},\ and\ \citenamefont {Saad}}]{Ahmed:2023pjl}%
  \BibitemOpen
  \bibfield  {author} {\bibinfo {author} {\bibfnamefont {Waqas}\ \bibnamefont {Ahmed}}, \bibinfo {author} {\bibfnamefont {Talal~Ahmed}\ \bibnamefont {Chowdhury}}, \bibinfo {author} {\bibfnamefont {Salah}\ \bibnamefont {Nasri}}, \ and\ \bibinfo {author} {\bibfnamefont {Shaikh}\ \bibnamefont {Saad}},\ }\bibfield  {title} {\enquote {\bibinfo {title} {{Gravitational waves from metastable cosmic strings in the Pati-Salam model in light of new pulsar timing array data}},}\ }\href {\doibase 10.1103/PhysRevD.109.015008} {\bibfield  {journal} {\bibinfo  {journal} {Phys. Rev. D}\ }\textbf {\bibinfo {volume} {109}},\ \bibinfo {pages} {015008} (\bibinfo {year} {2024})},\ \Eprint {http://arxiv.org/abs/2308.13248} {arXiv:2308.13248 [hep-ph]} \BibitemShut {NoStop}%
\bibitem [{\citenamefont {Antusch}\ \emph {et~al.}(2023)\citenamefont {Antusch}, \citenamefont {Hinze}, \citenamefont {Saad},\ and\ \citenamefont {Steiner}}]{Antusch:2023zjk}%
  \BibitemOpen
  \bibfield  {author} {\bibinfo {author} {\bibfnamefont {Stefan}\ \bibnamefont {Antusch}}, \bibinfo {author} {\bibfnamefont {Kevin}\ \bibnamefont {Hinze}}, \bibinfo {author} {\bibfnamefont {Shaikh}\ \bibnamefont {Saad}}, \ and\ \bibinfo {author} {\bibfnamefont {Jonathan}\ \bibnamefont {Steiner}},\ }\bibfield  {title} {\enquote {\bibinfo {title} {{Singling out SO(10) GUT models using recent PTA results}},}\ }\href {\doibase 10.1103/PhysRevD.108.095053} {\bibfield  {journal} {\bibinfo  {journal} {Phys. Rev. D}\ }\textbf {\bibinfo {volume} {108}},\ \bibinfo {pages} {095053} (\bibinfo {year} {2023})},\ \Eprint {http://arxiv.org/abs/2307.04595} {arXiv:2307.04595 [hep-ph]} \BibitemShut {NoStop}%
\bibitem [{\citenamefont {Ye}\ \emph {et~al.}(2024)\citenamefont {Ye}, \citenamefont {Zhu},\ and\ \citenamefont {Cai}}]{Ye:2023tpz}%
  \BibitemOpen
  \bibfield  {author} {\bibinfo {author} {\bibfnamefont {Gen}\ \bibnamefont {Ye}}, \bibinfo {author} {\bibfnamefont {Mian}\ \bibnamefont {Zhu}}, \ and\ \bibinfo {author} {\bibfnamefont {Yong}\ \bibnamefont {Cai}},\ }\bibfield  {title} {\enquote {\bibinfo {title} {{Null energy condition violation during inflation and pulsar timing array observations}},}\ }\href {\doibase 10.1007/JHEP02(2024)008} {\bibfield  {journal} {\bibinfo  {journal} {JHEP}\ }\textbf {\bibinfo {volume} {02}},\ \bibinfo {pages} {008} (\bibinfo {year} {2024})},\ \Eprint {http://arxiv.org/abs/2312.10685} {arXiv:2312.10685 [gr-qc]} \BibitemShut {NoStop}%
\bibitem [{\citenamefont {Cai}\ and\ \citenamefont {Piao}(2017)}]{Cai:2017dyi}%
  \BibitemOpen
  \bibfield  {author} {\bibinfo {author} {\bibfnamefont {Yong}\ \bibnamefont {Cai}}\ and\ \bibinfo {author} {\bibfnamefont {Yun-Song}\ \bibnamefont {Piao}},\ }\bibfield  {title} {\enquote {\bibinfo {title} {{A covariant Lagrangian for stable nonsingular bounce}},}\ }\href {\doibase 10.1007/JHEP09(2017)027} {\bibfield  {journal} {\bibinfo  {journal} {JHEP}\ }\textbf {\bibinfo {volume} {09}},\ \bibinfo {pages} {027} (\bibinfo {year} {2017})},\ \Eprint {http://arxiv.org/abs/1705.03401} {arXiv:1705.03401 [gr-qc]} \BibitemShut {NoStop}%
\bibitem [{\citenamefont {Kolevatov}\ \emph {et~al.}(2017)\citenamefont {Kolevatov}, \citenamefont {Mironov}, \citenamefont {Sukhov},\ and\ \citenamefont {Volkova}}]{Kolevatov:2017voe}%
  \BibitemOpen
  \bibfield  {author} {\bibinfo {author} {\bibfnamefont {R.}~\bibnamefont {Kolevatov}}, \bibinfo {author} {\bibfnamefont {S.}~\bibnamefont {Mironov}}, \bibinfo {author} {\bibfnamefont {N.}~\bibnamefont {Sukhov}}, \ and\ \bibinfo {author} {\bibfnamefont {V.}~\bibnamefont {Volkova}},\ }\bibfield  {title} {\enquote {\bibinfo {title} {{Cosmological bounce and Genesis beyond Horndeski}},}\ }\href {\doibase 10.1088/1475-7516/2017/08/038} {\bibfield  {journal} {\bibinfo  {journal} {JCAP}\ }\textbf {\bibinfo {volume} {08}},\ \bibinfo {pages} {038} (\bibinfo {year} {2017})},\ \Eprint {http://arxiv.org/abs/1705.06626} {arXiv:1705.06626 [hep-th]} \BibitemShut {NoStop}%
\bibitem [{\citenamefont {Ilyas}\ \emph {et~al.}(2021)\citenamefont {Ilyas}, \citenamefont {Zhu}, \citenamefont {Zheng},\ and\ \citenamefont {Cai}}]{Ilyas:2020zcb}%
  \BibitemOpen
  \bibfield  {author} {\bibinfo {author} {\bibfnamefont {Amara}\ \bibnamefont {Ilyas}}, \bibinfo {author} {\bibfnamefont {Mian}\ \bibnamefont {Zhu}}, \bibinfo {author} {\bibfnamefont {Yunlong}\ \bibnamefont {Zheng}}, \ and\ \bibinfo {author} {\bibfnamefont {Yi-Fu}\ \bibnamefont {Cai}},\ }\bibfield  {title} {\enquote {\bibinfo {title} {{Emergent Universe and Genesis from the DHOST Cosmology}},}\ }\href {\doibase 10.1007/JHEP01(2021)141} {\bibfield  {journal} {\bibinfo  {journal} {JHEP}\ }\textbf {\bibinfo {volume} {01}},\ \bibinfo {pages} {141} (\bibinfo {year} {2021})},\ \Eprint {http://arxiv.org/abs/2009.10351} {arXiv:2009.10351 [gr-qc]} \BibitemShut {NoStop}%
\bibitem [{\citenamefont {Ilyas}\ \emph {et~al.}(2020)\citenamefont {Ilyas}, \citenamefont {Zhu}, \citenamefont {Zheng}, \citenamefont {Cai},\ and\ \citenamefont {Saridakis}}]{Ilyas:2020qja}%
  \BibitemOpen
  \bibfield  {author} {\bibinfo {author} {\bibfnamefont {Amara}\ \bibnamefont {Ilyas}}, \bibinfo {author} {\bibfnamefont {Mian}\ \bibnamefont {Zhu}}, \bibinfo {author} {\bibfnamefont {Yunlong}\ \bibnamefont {Zheng}}, \bibinfo {author} {\bibfnamefont {Yi-Fu}\ \bibnamefont {Cai}}, \ and\ \bibinfo {author} {\bibfnamefont {Emmanuel~N.}\ \bibnamefont {Saridakis}},\ }\bibfield  {title} {\enquote {\bibinfo {title} {{DHOST Bounce}},}\ }\href {\doibase 10.1088/1475-7516/2020/09/002} {\bibfield  {journal} {\bibinfo  {journal} {JCAP}\ }\textbf {\bibinfo {volume} {09}},\ \bibinfo {pages} {002} (\bibinfo {year} {2020})},\ \Eprint {http://arxiv.org/abs/2002.08269} {arXiv:2002.08269 [gr-qc]} \BibitemShut {NoStop}%
\bibitem [{\citenamefont {Zhu}\ and\ \citenamefont {Zheng}(2021)}]{Zhu:2021ggm}%
  \BibitemOpen
  \bibfield  {author} {\bibinfo {author} {\bibfnamefont {Mian}\ \bibnamefont {Zhu}}\ and\ \bibinfo {author} {\bibfnamefont {Yunlong}\ \bibnamefont {Zheng}},\ }\bibfield  {title} {\enquote {\bibinfo {title} {{Improved DHOST Genesis}},}\ }\href {\doibase 10.1007/JHEP11(2021)163} {\bibfield  {journal} {\bibinfo  {journal} {JHEP}\ }\textbf {\bibinfo {volume} {11}},\ \bibinfo {pages} {163} (\bibinfo {year} {2021})},\ \Eprint {http://arxiv.org/abs/2109.05277} {arXiv:2109.05277 [gr-qc]} \BibitemShut {NoStop}%
\bibitem [{\citenamefont {Lesnefsky}\ \emph {et~al.}(2023)\citenamefont {Lesnefsky}, \citenamefont {Easson},\ and\ \citenamefont {Davies}}]{Lesnefsky:2022fen}%
  \BibitemOpen
  \bibfield  {author} {\bibinfo {author} {\bibfnamefont {J.~E.}\ \bibnamefont {Lesnefsky}}, \bibinfo {author} {\bibfnamefont {D.~A.}\ \bibnamefont {Easson}}, \ and\ \bibinfo {author} {\bibfnamefont {P.~C.~W.}\ \bibnamefont {Davies}},\ }\bibfield  {title} {\enquote {\bibinfo {title} {{Past-completeness of inflationary spacetimes}},}\ }\href {\doibase 10.1103/PhysRevD.107.044024} {\bibfield  {journal} {\bibinfo  {journal} {Phys. Rev. D}\ }\textbf {\bibinfo {volume} {107}},\ \bibinfo {pages} {044024} (\bibinfo {year} {2023})},\ \Eprint {http://arxiv.org/abs/2207.00955} {arXiv:2207.00955 [gr-qc]} \BibitemShut {NoStop}%
\bibitem [{\citenamefont {Cai}\ \emph {et~al.}(2024)\citenamefont {Cai}, \citenamefont {Zhu},\ and\ \citenamefont {Piao}}]{Cai:2023uhc}%
  \BibitemOpen
  \bibfield  {author} {\bibinfo {author} {\bibfnamefont {Yong}\ \bibnamefont {Cai}}, \bibinfo {author} {\bibfnamefont {Mian}\ \bibnamefont {Zhu}}, \ and\ \bibinfo {author} {\bibfnamefont {Yun-Song}\ \bibnamefont {Piao}},\ }\bibfield  {title} {\enquote {\bibinfo {title} {{Primordial Black Holes from Null Energy Condition Violation during Inflation}},}\ }\href {\doibase 10.1103/PhysRevLett.133.021001} {\bibfield  {journal} {\bibinfo  {journal} {Phys. Rev. Lett.}\ }\textbf {\bibinfo {volume} {133}},\ \bibinfo {pages} {021001} (\bibinfo {year} {2024})},\ \Eprint {http://arxiv.org/abs/2305.10933} {arXiv:2305.10933 [gr-qc]} \BibitemShut {NoStop}%
\bibitem [{\citenamefont {Chen}\ and\ \citenamefont {Liu}(2024{\natexlab{b}})}]{Chen:2024mwg}%
  \BibitemOpen
  \bibfield  {author} {\bibinfo {author} {\bibfnamefont {Zu-Cheng}\ \bibnamefont {Chen}}\ and\ \bibinfo {author} {\bibfnamefont {Lang}\ \bibnamefont {Liu}},\ }\bibfield  {title} {\enquote {\bibinfo {title} {{Constraints on inflation with null energy condition violation from advanced LIGO and advanced Virgo's first three observing runs}},}\ }\href {\doibase 10.1088/1475-7516/2024/06/028} {\bibfield  {journal} {\bibinfo  {journal} {JCAP}\ }\textbf {\bibinfo {volume} {06}},\ \bibinfo {pages} {028} (\bibinfo {year} {2024}{\natexlab{b}})},\ \Eprint {http://arxiv.org/abs/2404.07075} {arXiv:2404.07075 [gr-qc]} \BibitemShut {NoStop}%
\bibitem [{\citenamefont {Easson}\ and\ \citenamefont {Lesnefsky}(2024)}]{Easson:2024fzn}%
  \BibitemOpen
  \bibfield  {author} {\bibinfo {author} {\bibfnamefont {Damien~A.}\ \bibnamefont {Easson}}\ and\ \bibinfo {author} {\bibfnamefont {Joseph~E.}\ \bibnamefont {Lesnefsky}},\ }\bibfield  {title} {\enquote {\bibinfo {title} {{Eternal Universes}},}\ }\href@noop {} {\  (\bibinfo {year} {2024})},\ \Eprint {http://arxiv.org/abs/2404.03016} {arXiv:2404.03016 [hep-th]} \BibitemShut {NoStop}%
\bibitem [{\citenamefont {Rubakov}(2014)}]{Rubakov:2014jja}%
  \BibitemOpen
  \bibfield  {author} {\bibinfo {author} {\bibfnamefont {V.~A.}\ \bibnamefont {Rubakov}},\ }\bibfield  {title} {\enquote {\bibinfo {title} {{The Null Energy Condition and its violation}},}\ }\href {\doibase 10.3367/UFNe.0184.201402b.0137} {\bibfield  {journal} {\bibinfo  {journal} {Phys. Usp.}\ }\textbf {\bibinfo {volume} {57}},\ \bibinfo {pages} {128--142} (\bibinfo {year} {2014})},\ \Eprint {http://arxiv.org/abs/1401.4024} {arXiv:1401.4024 [hep-th]} \BibitemShut {NoStop}%
\bibitem [{\citenamefont {Cai}\ \emph {et~al.}(2012)\citenamefont {Cai}, \citenamefont {Easson},\ and\ \citenamefont {Brandenberger}}]{Cai:2012va}%
  \BibitemOpen
  \bibfield  {author} {\bibinfo {author} {\bibfnamefont {Yi-Fu}\ \bibnamefont {Cai}}, \bibinfo {author} {\bibfnamefont {Damien~A.}\ \bibnamefont {Easson}}, \ and\ \bibinfo {author} {\bibfnamefont {Robert}\ \bibnamefont {Brandenberger}},\ }\bibfield  {title} {\enquote {\bibinfo {title} {{Towards a Nonsingular Bouncing Cosmology}},}\ }\href {\doibase 10.1088/1475-7516/2012/08/020} {\bibfield  {journal} {\bibinfo  {journal} {JCAP}\ }\textbf {\bibinfo {volume} {08}},\ \bibinfo {pages} {020} (\bibinfo {year} {2012})},\ \Eprint {http://arxiv.org/abs/1206.2382} {arXiv:1206.2382 [hep-th]} \BibitemShut {NoStop}%
\bibitem [{\citenamefont {Libanov}\ \emph {et~al.}(2016)\citenamefont {Libanov}, \citenamefont {Mironov},\ and\ \citenamefont {Rubakov}}]{Libanov:2016kfc}%
  \BibitemOpen
  \bibfield  {author} {\bibinfo {author} {\bibfnamefont {M.}~\bibnamefont {Libanov}}, \bibinfo {author} {\bibfnamefont {S.}~\bibnamefont {Mironov}}, \ and\ \bibinfo {author} {\bibfnamefont {V.}~\bibnamefont {Rubakov}},\ }\bibfield  {title} {\enquote {\bibinfo {title} {{Generalized Galileons: instabilities of bouncing and Genesis cosmologies and modified Genesis}},}\ }\href {\doibase 10.1088/1475-7516/2016/08/037} {\bibfield  {journal} {\bibinfo  {journal} {JCAP}\ }\textbf {\bibinfo {volume} {08}},\ \bibinfo {pages} {037} (\bibinfo {year} {2016})},\ \Eprint {http://arxiv.org/abs/1605.05992} {arXiv:1605.05992 [hep-th]} \BibitemShut {NoStop}%
\bibitem [{\citenamefont {Kobayashi}(2016)}]{Kobayashi:2016xpl}%
  \BibitemOpen
  \bibfield  {author} {\bibinfo {author} {\bibfnamefont {Tsutomu}\ \bibnamefont {Kobayashi}},\ }\bibfield  {title} {\enquote {\bibinfo {title} {{Generic instabilities of nonsingular cosmologies in Horndeski theory: A no-go theorem}},}\ }\href {\doibase 10.1103/PhysRevD.94.043511} {\bibfield  {journal} {\bibinfo  {journal} {Phys. Rev. D}\ }\textbf {\bibinfo {volume} {94}},\ \bibinfo {pages} {043511} (\bibinfo {year} {2016})},\ \Eprint {http://arxiv.org/abs/1606.05831} {arXiv:1606.05831 [hep-th]} \BibitemShut {NoStop}%
\bibitem [{\citenamefont {Dobre}\ \emph {et~al.}(2018)\citenamefont {Dobre}, \citenamefont {Frolov}, \citenamefont {G\'alvez~Ghersi}, \citenamefont {Ramazanov},\ and\ \citenamefont {Vikman}}]{Dobre:2017pnt}%
  \BibitemOpen
  \bibfield  {author} {\bibinfo {author} {\bibfnamefont {David~A.}\ \bibnamefont {Dobre}}, \bibinfo {author} {\bibfnamefont {Andrei~V.}\ \bibnamefont {Frolov}}, \bibinfo {author} {\bibfnamefont {Jos\'e~T.}\ \bibnamefont {G\'alvez~Ghersi}}, \bibinfo {author} {\bibfnamefont {Sabir}\ \bibnamefont {Ramazanov}}, \ and\ \bibinfo {author} {\bibfnamefont {Alexander}\ \bibnamefont {Vikman}},\ }\bibfield  {title} {\enquote {\bibinfo {title} {{Unbraiding the Bounce: Superluminality around the Corner}},}\ }\href {\doibase 10.1088/1475-7516/2018/03/020} {\bibfield  {journal} {\bibinfo  {journal} {JCAP}\ }\textbf {\bibinfo {volume} {03}},\ \bibinfo {pages} {020} (\bibinfo {year} {2018})},\ \Eprint {http://arxiv.org/abs/1712.10272} {arXiv:1712.10272 [gr-qc]} \BibitemShut {NoStop}%
\bibitem [{\citenamefont {Cai}\ \emph {et~al.}(2022)\citenamefont {Cai}, \citenamefont {Xu}, \citenamefont {Zhao},\ and\ \citenamefont {Zhou}}]{Cai:2022ori}%
  \BibitemOpen
  \bibfield  {author} {\bibinfo {author} {\bibfnamefont {Yong}\ \bibnamefont {Cai}}, \bibinfo {author} {\bibfnamefont {Ji}~\bibnamefont {Xu}}, \bibinfo {author} {\bibfnamefont {Shuai}\ \bibnamefont {Zhao}}, \ and\ \bibinfo {author} {\bibfnamefont {Siyi}\ \bibnamefont {Zhou}},\ }\bibfield  {title} {\enquote {\bibinfo {title} {{Perturbative unitarity and NEC violation in genesis cosmology}},}\ }\href {\doibase 10.1007/JHEP10(2022)140} {\bibfield  {journal} {\bibinfo  {journal} {JHEP}\ }\textbf {\bibinfo {volume} {10}},\ \bibinfo {pages} {140} (\bibinfo {year} {2022})},\ \bibinfo {note} {[Erratum: JHEP 11, 063 (2022)]},\ \Eprint {http://arxiv.org/abs/2207.11772} {arXiv:2207.11772 [gr-qc]} \BibitemShut {NoStop}%
\bibitem [{\citenamefont {Cai}\ and\ \citenamefont {Piao}(2021)}]{Cai:2020qpu}%
  \BibitemOpen
  \bibfield  {author} {\bibinfo {author} {\bibfnamefont {Yong}\ \bibnamefont {Cai}}\ and\ \bibinfo {author} {\bibfnamefont {Yun-Song}\ \bibnamefont {Piao}},\ }\bibfield  {title} {\enquote {\bibinfo {title} {{Intermittent null energy condition violations during inflation and primordial gravitational waves}},}\ }\href {\doibase 10.1103/PhysRevD.103.083521} {\bibfield  {journal} {\bibinfo  {journal} {Phys. Rev. D}\ }\textbf {\bibinfo {volume} {103}},\ \bibinfo {pages} {083521} (\bibinfo {year} {2021})},\ \Eprint {http://arxiv.org/abs/2012.11304} {arXiv:2012.11304 [gr-qc]} \BibitemShut {NoStop}%
\bibitem [{\citenamefont {Cai}\ and\ \citenamefont {Piao}(2022)}]{Cai:2022nqv}%
  \BibitemOpen
  \bibfield  {author} {\bibinfo {author} {\bibfnamefont {Yong}\ \bibnamefont {Cai}}\ and\ \bibinfo {author} {\bibfnamefont {Yun-Song}\ \bibnamefont {Piao}},\ }\bibfield  {title} {\enquote {\bibinfo {title} {{Generating enhanced primordial GWs during inflation with intermittent violation of NEC and diminishment of GW propagating speed}},}\ }\href {\doibase 10.1007/JHEP06(2022)067} {\bibfield  {journal} {\bibinfo  {journal} {JHEP}\ }\textbf {\bibinfo {volume} {06}},\ \bibinfo {pages} {067} (\bibinfo {year} {2022})},\ \Eprint {http://arxiv.org/abs/2201.04552} {arXiv:2201.04552 [gr-qc]} \BibitemShut {NoStop}%
\bibitem [{\citenamefont {Ruan}\ \emph {et~al.}(2020)\citenamefont {Ruan}, \citenamefont {Guo}, \citenamefont {Cai},\ and\ \citenamefont {Zhang}}]{Ruan:2018tsw}%
  \BibitemOpen
  \bibfield  {author} {\bibinfo {author} {\bibfnamefont {Wen-Hong}\ \bibnamefont {Ruan}}, \bibinfo {author} {\bibfnamefont {Zong-Kuan}\ \bibnamefont {Guo}}, \bibinfo {author} {\bibfnamefont {Rong-Gen}\ \bibnamefont {Cai}}, \ and\ \bibinfo {author} {\bibfnamefont {Yuan-Zhong}\ \bibnamefont {Zhang}},\ }\bibfield  {title} {\enquote {\bibinfo {title} {{Taiji program: Gravitational-wave sources}},}\ }\href {\doibase 10.1142/S0217751X2050075X} {\bibfield  {journal} {\bibinfo  {journal} {Int. J. Mod. Phys. A}\ }\textbf {\bibinfo {volume} {35}},\ \bibinfo {pages} {2050075} (\bibinfo {year} {2020})},\ \Eprint {http://arxiv.org/abs/1807.09495} {arXiv:1807.09495 [gr-qc]} \BibitemShut {NoStop}%
\bibitem [{\citenamefont {Turner}\ \emph {et~al.}(1993)\citenamefont {Turner}, \citenamefont {White},\ and\ \citenamefont {Lidsey}}]{Turner:1993vb}%
  \BibitemOpen
  \bibfield  {author} {\bibinfo {author} {\bibfnamefont {Michael~S.}\ \bibnamefont {Turner}}, \bibinfo {author} {\bibfnamefont {Martin~J.}\ \bibnamefont {White}}, \ and\ \bibinfo {author} {\bibfnamefont {James~E.}\ \bibnamefont {Lidsey}},\ }\bibfield  {title} {\enquote {\bibinfo {title} {{Tensor perturbations in inflationary models as a probe of cosmology}},}\ }\href {\doibase 10.1103/PhysRevD.48.4613} {\bibfield  {journal} {\bibinfo  {journal} {Phys. Rev. D}\ }\textbf {\bibinfo {volume} {48}},\ \bibinfo {pages} {4613--4622} (\bibinfo {year} {1993})},\ \Eprint {http://arxiv.org/abs/astro-ph/9306029} {arXiv:astro-ph/9306029} \BibitemShut {NoStop}%
\bibitem [{\citenamefont {Aghanim}\ \emph {et~al.}(2020)\citenamefont {Aghanim} \emph {et~al.}}]{Planck:2018vyg}%
  \BibitemOpen
  \bibfield  {author} {\bibinfo {author} {\bibfnamefont {N.}~\bibnamefont {Aghanim}} \emph {et~al.} (\bibinfo {collaboration} {Planck}),\ }\bibfield  {title} {\enquote {\bibinfo {title} {{Planck 2018 results. VI. Cosmological parameters}},}\ }\href {\doibase 10.1051/0004-6361/201833910} {\bibfield  {journal} {\bibinfo  {journal} {Astron. Astrophys.}\ }\textbf {\bibinfo {volume} {641}},\ \bibinfo {pages} {A6} (\bibinfo {year} {2020})},\ \bibinfo {note} {[Erratum: Astron.Astrophys. 652, C4 (2021)]},\ \Eprint {http://arxiv.org/abs/1807.06209} {arXiv:1807.06209 [astro-ph.CO]} \BibitemShut {NoStop}%
\bibitem [{\citenamefont {Wu}\ \emph {et~al.}(2024)\citenamefont {Wu}, \citenamefont {Chen},\ and\ \citenamefont {Huang}}]{Wu:2023hsa}%
  \BibitemOpen
  \bibfield  {author} {\bibinfo {author} {\bibfnamefont {Yu-Mei}\ \bibnamefont {Wu}}, \bibinfo {author} {\bibfnamefont {Zu-Cheng}\ \bibnamefont {Chen}}, \ and\ \bibinfo {author} {\bibfnamefont {Qing-Guo}\ \bibnamefont {Huang}},\ }\bibfield  {title} {\enquote {\bibinfo {title} {{Cosmological interpretation for the stochastic signal in pulsar timing arrays}},}\ }\href {\doibase 10.1007/s11433-023-2298-7} {\bibfield  {journal} {\bibinfo  {journal} {Sci. China Phys. Mech. Astron.}\ }\textbf {\bibinfo {volume} {67}},\ \bibinfo {pages} {240412} (\bibinfo {year} {2024})},\ \Eprint {http://arxiv.org/abs/2307.03141} {arXiv:2307.03141 [astro-ph.CO]} \BibitemShut {NoStop}%
\bibitem [{\citenamefont {Meerburg}\ \emph {et~al.}(2015)\citenamefont {Meerburg}, \citenamefont {Hlo\v{z}ek}, \citenamefont {Hadzhiyska},\ and\ \citenamefont {Meyers}}]{Meerburg:2015zua}%
  \BibitemOpen
  \bibfield  {author} {\bibinfo {author} {\bibfnamefont {P.~Daniel}\ \bibnamefont {Meerburg}}, \bibinfo {author} {\bibfnamefont {Ren\'ee}\ \bibnamefont {Hlo\v{z}ek}}, \bibinfo {author} {\bibfnamefont {Boryana}\ \bibnamefont {Hadzhiyska}}, \ and\ \bibinfo {author} {\bibfnamefont {Joel}\ \bibnamefont {Meyers}},\ }\bibfield  {title} {\enquote {\bibinfo {title} {{Multiwavelength constraints on the inflationary consistency relation}},}\ }\href {\doibase 10.1103/PhysRevD.91.103505} {\bibfield  {journal} {\bibinfo  {journal} {Phys. Rev. D}\ }\textbf {\bibinfo {volume} {91}},\ \bibinfo {pages} {103505} (\bibinfo {year} {2015})},\ \Eprint {http://arxiv.org/abs/1502.00302} {arXiv:1502.00302 [astro-ph.CO]} \BibitemShut {NoStop}%
\bibitem [{\citenamefont {Ben-Dayan}\ \emph {et~al.}(2019)\citenamefont {Ben-Dayan}, \citenamefont {Keating}, \citenamefont {Leon},\ and\ \citenamefont {Wolfson}}]{Ben-Dayan:2019gll}%
  \BibitemOpen
  \bibfield  {author} {\bibinfo {author} {\bibfnamefont {Ido}\ \bibnamefont {Ben-Dayan}}, \bibinfo {author} {\bibfnamefont {Brian}\ \bibnamefont {Keating}}, \bibinfo {author} {\bibfnamefont {David}\ \bibnamefont {Leon}}, \ and\ \bibinfo {author} {\bibfnamefont {Ira}\ \bibnamefont {Wolfson}},\ }\bibfield  {title} {\enquote {\bibinfo {title} {{Constraints on scalar and tensor spectra from $N_{eff}$}},}\ }\href {\doibase 10.1088/1475-7516/2019/06/007} {\bibfield  {journal} {\bibinfo  {journal} {JCAP}\ }\textbf {\bibinfo {volume} {06}},\ \bibinfo {pages} {007} (\bibinfo {year} {2019})},\ \Eprint {http://arxiv.org/abs/1903.11843} {arXiv:1903.11843 [astro-ph.CO]} \BibitemShut {NoStop}%
\bibitem [{\citenamefont {Cabass}\ \emph {et~al.}(2016)\citenamefont {Cabass}, \citenamefont {Pagano}, \citenamefont {Salvati}, \citenamefont {Gerbino}, \citenamefont {Giusarma},\ and\ \citenamefont {Melchiorri}}]{Cabass:2015jwe}%
  \BibitemOpen
  \bibfield  {author} {\bibinfo {author} {\bibfnamefont {Giovanni}\ \bibnamefont {Cabass}}, \bibinfo {author} {\bibfnamefont {Luca}\ \bibnamefont {Pagano}}, \bibinfo {author} {\bibfnamefont {Laura}\ \bibnamefont {Salvati}}, \bibinfo {author} {\bibfnamefont {Martina}\ \bibnamefont {Gerbino}}, \bibinfo {author} {\bibfnamefont {Elena}\ \bibnamefont {Giusarma}}, \ and\ \bibinfo {author} {\bibfnamefont {Alessandro}\ \bibnamefont {Melchiorri}},\ }\bibfield  {title} {\enquote {\bibinfo {title} {{Updated Constraints and Forecasts on Primordial Tensor Modes}},}\ }\href {\doibase 10.1103/PhysRevD.93.063508} {\bibfield  {journal} {\bibinfo  {journal} {Phys. Rev. D}\ }\textbf {\bibinfo {volume} {93}},\ \bibinfo {pages} {063508} (\bibinfo {year} {2016})},\ \Eprint {http://arxiv.org/abs/1511.05146} {arXiv:1511.05146 [astro-ph.CO]} \BibitemShut {NoStop}%
\bibitem [{\citenamefont {Abbott}\ \emph {et~al.}(2021{\natexlab{d}})\citenamefont {Abbott} \emph {et~al.}}]{KAGRA:2021kbb}%
  \BibitemOpen
  \bibfield  {author} {\bibinfo {author} {\bibfnamefont {R.}~\bibnamefont {Abbott}} \emph {et~al.} (\bibinfo {collaboration} {KAGRA, Virgo, LIGO Scientific}),\ }\bibfield  {title} {\enquote {\bibinfo {title} {{Upper limits on the isotropic gravitational-wave background from Advanced LIGO and Advanced Virgo\textquoteright{}s third observing run}},}\ }\href {\doibase 10.1103/PhysRevD.104.022004} {\bibfield  {journal} {\bibinfo  {journal} {Phys. Rev. D}\ }\textbf {\bibinfo {volume} {104}},\ \bibinfo {pages} {022004} (\bibinfo {year} {2021}{\natexlab{d}})},\ \Eprint {http://arxiv.org/abs/2101.12130} {arXiv:2101.12130 [gr-qc]} \BibitemShut {NoStop}%
\bibitem [{\citenamefont {Tinto}\ \emph {et~al.}(2001)\citenamefont {Tinto}, \citenamefont {Armstrong},\ and\ \citenamefont {Estabrook}}]{Tinto:2001ii}%
  \BibitemOpen
  \bibfield  {author} {\bibinfo {author} {\bibfnamefont {Massimo}\ \bibnamefont {Tinto}}, \bibinfo {author} {\bibfnamefont {J.~W.}\ \bibnamefont {Armstrong}}, \ and\ \bibinfo {author} {\bibfnamefont {F.~B.}\ \bibnamefont {Estabrook}},\ }\bibfield  {title} {\enquote {\bibinfo {title} {{Discriminating a gravitational wave background from instrumental noise in the LISA detector}},}\ }\href {\doibase 10.1103/PhysRevD.63.021101} {\bibfield  {journal} {\bibinfo  {journal} {Phys. Rev. D}\ }\textbf {\bibinfo {volume} {63}},\ \bibinfo {pages} {021101} (\bibinfo {year} {2001})}\BibitemShut {NoStop}%
\bibitem [{\citenamefont {Tinto}\ \emph {et~al.}(2002)\citenamefont {Tinto}, \citenamefont {Estabrook},\ and\ \citenamefont {Armstrong}}]{Tinto:2002de}%
  \BibitemOpen
  \bibfield  {author} {\bibinfo {author} {\bibfnamefont {Massimo}\ \bibnamefont {Tinto}}, \bibinfo {author} {\bibfnamefont {F.~B.}\ \bibnamefont {Estabrook}}, \ and\ \bibinfo {author} {\bibfnamefont {J.~W.}\ \bibnamefont {Armstrong}},\ }\bibfield  {title} {\enquote {\bibinfo {title} {{Time delay interferometry for LISA}},}\ }\href {\doibase 10.1103/PhysRevD.65.082003} {\bibfield  {journal} {\bibinfo  {journal} {Phys. Rev. D}\ }\textbf {\bibinfo {volume} {65}},\ \bibinfo {pages} {082003} (\bibinfo {year} {2002})}\BibitemShut {NoStop}%
\bibitem [{\citenamefont {Vallisneri}\ and\ \citenamefont {Galley}(2012)}]{Vallisneri:2012np}%
  \BibitemOpen
  \bibfield  {author} {\bibinfo {author} {\bibfnamefont {Michele}\ \bibnamefont {Vallisneri}}\ and\ \bibinfo {author} {\bibfnamefont {Chad~R.}\ \bibnamefont {Galley}},\ }\bibfield  {title} {\enquote {\bibinfo {title} {{Non-sky-averaged sensitivity curves for space-based gravitational-wave observatories}},}\ }\href {\doibase 10.1088/0264-9381/29/12/124015} {\bibfield  {journal} {\bibinfo  {journal} {Class. Quant. Grav.}\ }\textbf {\bibinfo {volume} {29}},\ \bibinfo {pages} {124015} (\bibinfo {year} {2012})},\ \Eprint {http://arxiv.org/abs/1201.3684} {arXiv:1201.3684 [gr-qc]} \BibitemShut {NoStop}%
\bibitem [{\citenamefont {Flauger}\ \emph {et~al.}(2021)\citenamefont {Flauger}, \citenamefont {Karnesis}, \citenamefont {Nardini}, \citenamefont {Pieroni}, \citenamefont {Ricciardone},\ and\ \citenamefont {Torrado}}]{Flauger:2020qyi}%
  \BibitemOpen
  \bibfield  {author} {\bibinfo {author} {\bibfnamefont {Raphael}\ \bibnamefont {Flauger}}, \bibinfo {author} {\bibfnamefont {Nikolaos}\ \bibnamefont {Karnesis}}, \bibinfo {author} {\bibfnamefont {Germano}\ \bibnamefont {Nardini}}, \bibinfo {author} {\bibfnamefont {Mauro}\ \bibnamefont {Pieroni}}, \bibinfo {author} {\bibfnamefont {Angelo}\ \bibnamefont {Ricciardone}}, \ and\ \bibinfo {author} {\bibfnamefont {Jes\'us}\ \bibnamefont {Torrado}},\ }\bibfield  {title} {\enquote {\bibinfo {title} {{Improved reconstruction of a stochastic gravitational wave background with LISA}},}\ }\href {\doibase 10.1088/1475-7516/2021/01/059} {\bibfield  {journal} {\bibinfo  {journal} {JCAP}\ }\textbf {\bibinfo {volume} {01}},\ \bibinfo {pages} {059} (\bibinfo {year} {2021})},\ \Eprint {http://arxiv.org/abs/2009.11845} {arXiv:2009.11845 [astro-ph.CO]} \BibitemShut {NoStop}%
\bibitem [{\citenamefont {Ren}\ \emph {et~al.}(2023)\citenamefont {Ren}, \citenamefont {Zhao}, \citenamefont {Cao}, \citenamefont {Guo}, \citenamefont {Han}, \citenamefont {Jin},\ and\ \citenamefont {Wu}}]{Ren:2023yec}%
  \BibitemOpen
  \bibfield  {author} {\bibinfo {author} {\bibfnamefont {Zhixiang}\ \bibnamefont {Ren}}, \bibinfo {author} {\bibfnamefont {Tianyu}\ \bibnamefont {Zhao}}, \bibinfo {author} {\bibfnamefont {Zhoujian}\ \bibnamefont {Cao}}, \bibinfo {author} {\bibfnamefont {Zong-Kuan}\ \bibnamefont {Guo}}, \bibinfo {author} {\bibfnamefont {Wen-Biao}\ \bibnamefont {Han}}, \bibinfo {author} {\bibfnamefont {Hong-Bo}\ \bibnamefont {Jin}}, \ and\ \bibinfo {author} {\bibfnamefont {Yue-Liang}\ \bibnamefont {Wu}},\ }\bibfield  {title} {\enquote {\bibinfo {title} {{Taiji data challenge for exploring gravitational wave universe}},}\ }\href {\doibase 10.1007/s11467-023-1318-y} {\bibfield  {journal} {\bibinfo  {journal} {Front. Phys. (Beijing)}\ }\textbf {\bibinfo {volume} {18}},\ \bibinfo {pages} {64302} (\bibinfo {year} {2023})},\ \Eprint {http://arxiv.org/abs/2301.02967} {arXiv:2301.02967 [gr-qc]} \BibitemShut {NoStop}%
\bibitem [{\citenamefont {Luo}\ \emph {et~al.}(2020)\citenamefont {Luo}, \citenamefont {Guo}, \citenamefont {Jin}, \citenamefont {Wu},\ and\ \citenamefont {Hu}}]{Luo:2019zal}%
  \BibitemOpen
  \bibfield  {author} {\bibinfo {author} {\bibfnamefont {Ziren}\ \bibnamefont {Luo}}, \bibinfo {author} {\bibfnamefont {ZongKuan}\ \bibnamefont {Guo}}, \bibinfo {author} {\bibfnamefont {Gang}\ \bibnamefont {Jin}}, \bibinfo {author} {\bibfnamefont {Yueliang}\ \bibnamefont {Wu}}, \ and\ \bibinfo {author} {\bibfnamefont {Wenrui}\ \bibnamefont {Hu}},\ }\bibfield  {title} {\enquote {\bibinfo {title} {{A brief analysis to Taiji: Science and technology}},}\ }\href {\doibase 10.1016/j.rinp.2019.102918} {\bibfield  {journal} {\bibinfo  {journal} {Results Phys.}\ }\textbf {\bibinfo {volume} {16}},\ \bibinfo {pages} {102918} (\bibinfo {year} {2020})}\BibitemShut {NoStop}%
\bibitem [{\citenamefont {Smith}\ \emph {et~al.}(2019)\citenamefont {Smith}, \citenamefont {Smith}, \citenamefont {Caldwell},\ and\ \citenamefont {Caldwell}}]{Smith:2019wny}%
  \BibitemOpen
  \bibfield  {author} {\bibinfo {author} {\bibfnamefont {Tristan~L.}\ \bibnamefont {Smith}}, \bibinfo {author} {\bibfnamefont {Tristan~L.}\ \bibnamefont {Smith}}, \bibinfo {author} {\bibfnamefont {Robert~R.}\ \bibnamefont {Caldwell}}, \ and\ \bibinfo {author} {\bibfnamefont {Robert}\ \bibnamefont {Caldwell}},\ }\bibfield  {title} {\enquote {\bibinfo {title} {{LISA for Cosmologists: Calculating the Signal-to-Noise Ratio for Stochastic and Deterministic Sources}},}\ }\href {\doibase 10.1103/PhysRevD.100.104055} {\bibfield  {journal} {\bibinfo  {journal} {Phys. Rev. D}\ }\textbf {\bibinfo {volume} {100}},\ \bibinfo {pages} {104055} (\bibinfo {year} {2019})},\ \bibinfo {note} {[Erratum: Phys.Rev.D 105, 029902 (2022)]},\ \Eprint {http://arxiv.org/abs/1908.00546} {arXiv:1908.00546 [astro-ph.CO]} \BibitemShut {NoStop}%
\bibitem [{\citenamefont {Wang}\ \emph {et~al.}(2021)\citenamefont {Wang}, \citenamefont {Tan}, \citenamefont {Qian},\ and\ \citenamefont {Shao}}]{Wang:2021owg}%
  \BibitemOpen
  \bibfield  {author} {\bibinfo {author} {\bibfnamefont {Pan-Pan}\ \bibnamefont {Wang}}, \bibinfo {author} {\bibfnamefont {Yu-Jie}\ \bibnamefont {Tan}}, \bibinfo {author} {\bibfnamefont {Wei-Liang}\ \bibnamefont {Qian}}, \ and\ \bibinfo {author} {\bibfnamefont {Cheng-Gang}\ \bibnamefont {Shao}},\ }\bibfield  {title} {\enquote {\bibinfo {title} {{Sensitivity functions of space-borne gravitational wave detectors for arbitrary time-delay interferometry combinations regarding nontensorial polarizations}},}\ }\href {\doibase 10.1103/PhysRevD.104.023002} {\bibfield  {journal} {\bibinfo  {journal} {Phys. Rev. D}\ }\textbf {\bibinfo {volume} {104}},\ \bibinfo {pages} {023002} (\bibinfo {year} {2021})}\BibitemShut {NoStop}%
\bibitem [{\citenamefont {Korol}\ \emph {et~al.}(2020)\citenamefont {Korol} \emph {et~al.}}]{Korol:2020lpq}%
  \BibitemOpen
  \bibfield  {author} {\bibinfo {author} {\bibfnamefont {V.}~\bibnamefont {Korol}} \emph {et~al.},\ }\bibfield  {title} {\enquote {\bibinfo {title} {{Populations of double white dwarfs in Milky Way satellites and their detectability with LISA}},}\ }\href {\doibase 10.1051/0004-6361/202037764} {\bibfield  {journal} {\bibinfo  {journal} {Astron. Astrophys.}\ }\textbf {\bibinfo {volume} {638}},\ \bibinfo {pages} {A153} (\bibinfo {year} {2020})},\ \Eprint {http://arxiv.org/abs/2002.10462} {arXiv:2002.10462 [astro-ph.GA]} \BibitemShut {NoStop}%
\bibitem [{\citenamefont {Korol}\ \emph {et~al.}(2022)\citenamefont {Korol}, \citenamefont {Hallakoun}, \citenamefont {Toonen},\ and\ \citenamefont {Karnesis}}]{Korol:2021pun}%
  \BibitemOpen
  \bibfield  {author} {\bibinfo {author} {\bibfnamefont {Valeriya}\ \bibnamefont {Korol}}, \bibinfo {author} {\bibfnamefont {Na'ama}\ \bibnamefont {Hallakoun}}, \bibinfo {author} {\bibfnamefont {Silvia}\ \bibnamefont {Toonen}}, \ and\ \bibinfo {author} {\bibfnamefont {Nikolaos}\ \bibnamefont {Karnesis}},\ }\bibfield  {title} {\enquote {\bibinfo {title} {{Observationally driven Galactic double white dwarf population for LISA}},}\ }\href {\doibase 10.1093/mnras/stac415} {\bibfield  {journal} {\bibinfo  {journal} {Mon. Not. Roy. Astron. Soc.}\ }\textbf {\bibinfo {volume} {511}},\ \bibinfo {pages} {5936--5947} (\bibinfo {year} {2022})},\ \Eprint {http://arxiv.org/abs/2109.10972} {arXiv:2109.10972 [astro-ph.HE]} \BibitemShut {NoStop}%
\bibitem [{\citenamefont {Karnesis}\ \emph {et~al.}(2021)\citenamefont {Karnesis}, \citenamefont {Babak}, \citenamefont {Pieroni}, \citenamefont {Cornish},\ and\ \citenamefont {Littenberg}}]{Karnesis:2021tsh}%
  \BibitemOpen
  \bibfield  {author} {\bibinfo {author} {\bibfnamefont {Nikolaos}\ \bibnamefont {Karnesis}}, \bibinfo {author} {\bibfnamefont {Stanislav}\ \bibnamefont {Babak}}, \bibinfo {author} {\bibfnamefont {Mauro}\ \bibnamefont {Pieroni}}, \bibinfo {author} {\bibfnamefont {Neil}\ \bibnamefont {Cornish}}, \ and\ \bibinfo {author} {\bibfnamefont {Tyson}\ \bibnamefont {Littenberg}},\ }\bibfield  {title} {\enquote {\bibinfo {title} {{Characterization of the stochastic signal originating from compact binary populations as measured by LISA}},}\ }\href {\doibase 10.1103/PhysRevD.104.043019} {\bibfield  {journal} {\bibinfo  {journal} {Phys. Rev. D}\ }\textbf {\bibinfo {volume} {104}},\ \bibinfo {pages} {043019} (\bibinfo {year} {2021})},\ \Eprint {http://arxiv.org/abs/2103.14598} {arXiv:2103.14598 [astro-ph.IM]} \BibitemShut {NoStop}%
\bibitem [{\citenamefont {Liu}\ \emph {et~al.}(2023{\natexlab{b}})\citenamefont {Liu}, \citenamefont {Ruan},\ and\ \citenamefont {Guo}}]{Liu:2023qap}%
  \BibitemOpen
  \bibfield  {author} {\bibinfo {author} {\bibfnamefont {Chang}\ \bibnamefont {Liu}}, \bibinfo {author} {\bibfnamefont {Wen-Hong}\ \bibnamefont {Ruan}}, \ and\ \bibinfo {author} {\bibfnamefont {Zong-Kuan}\ \bibnamefont {Guo}},\ }\bibfield  {title} {\enquote {\bibinfo {title} {{Confusion noise from Galactic binaries for Taiji}},}\ }\href {\doibase 10.1103/PhysRevD.107.064021} {\bibfield  {journal} {\bibinfo  {journal} {Phys. Rev. D}\ }\textbf {\bibinfo {volume} {107}},\ \bibinfo {pages} {064021} (\bibinfo {year} {2023}{\natexlab{b}})},\ \Eprint {http://arxiv.org/abs/2301.02821} {arXiv:2301.02821 [astro-ph.IM]} \BibitemShut {NoStop}%
\bibitem [{\citenamefont {Chen}\ \emph {et~al.}(2024{\natexlab{c}})\citenamefont {Chen}, \citenamefont {Huang}, \citenamefont {Liu}, \citenamefont {Liu}, \citenamefont {Liu}, \citenamefont {Wu}, \citenamefont {Wu}, \citenamefont {Yi},\ and\ \citenamefont {You}}]{Chen:2023zkb}%
  \BibitemOpen
  \bibfield  {author} {\bibinfo {author} {\bibfnamefont {Zu-Cheng}\ \bibnamefont {Chen}}, \bibinfo {author} {\bibfnamefont {Qing-Guo}\ \bibnamefont {Huang}}, \bibinfo {author} {\bibfnamefont {Chang}\ \bibnamefont {Liu}}, \bibinfo {author} {\bibfnamefont {Lang}\ \bibnamefont {Liu}}, \bibinfo {author} {\bibfnamefont {Xiao-Jin}\ \bibnamefont {Liu}}, \bibinfo {author} {\bibfnamefont {You}\ \bibnamefont {Wu}}, \bibinfo {author} {\bibfnamefont {Yu-Mei}\ \bibnamefont {Wu}}, \bibinfo {author} {\bibfnamefont {Zhu}\ \bibnamefont {Yi}}, \ and\ \bibinfo {author} {\bibfnamefont {Zhi-Qiang}\ \bibnamefont {You}},\ }\bibfield  {title} {\enquote {\bibinfo {title} {{Prospects for Taiji to detect a gravitational-wave background from cosmic strings}},}\ }\href {\doibase 10.1088/1475-7516/2024/03/022} {\bibfield  {journal} {\bibinfo  {journal} {JCAP}\ }\textbf {\bibinfo {volume} {03}},\ \bibinfo {pages} {022} (\bibinfo {year} {2024}{\natexlab{c}})},\ \Eprint {http://arxiv.org/abs/2310.00411} {arXiv:2310.00411 [astro-ph.IM]}
  \BibitemShut {NoStop}%
\bibitem [{\citenamefont {Chen}\ \emph {et~al.}(2019)\citenamefont {Chen}, \citenamefont {Huang},\ and\ \citenamefont {Huang}}]{Chen:2018rzo}%
  \BibitemOpen
  \bibfield  {author} {\bibinfo {author} {\bibfnamefont {Zu-Cheng}\ \bibnamefont {Chen}}, \bibinfo {author} {\bibfnamefont {Fan}\ \bibnamefont {Huang}}, \ and\ \bibinfo {author} {\bibfnamefont {Qing-Guo}\ \bibnamefont {Huang}},\ }\bibfield  {title} {\enquote {\bibinfo {title} {{Stochastic Gravitational-wave Background from Binary Black Holes and Binary Neutron Stars and Implications for LISA}},}\ }\href {\doibase 10.3847/1538-4357/aaf581} {\bibfield  {journal} {\bibinfo  {journal} {Astrophys. J.}\ }\textbf {\bibinfo {volume} {871}},\ \bibinfo {pages} {97} (\bibinfo {year} {2019})},\ \Eprint {http://arxiv.org/abs/1809.10360} {arXiv:1809.10360 [gr-qc]} \BibitemShut {NoStop}%
\bibitem [{\citenamefont {Caprini}\ \emph {et~al.}(2019)\citenamefont {Caprini}, \citenamefont {Figueroa}, \citenamefont {Flauger}, \citenamefont {Nardini}, \citenamefont {Peloso}, \citenamefont {Pieroni}, \citenamefont {Ricciardone},\ and\ \citenamefont {Tasinato}}]{Caprini:2019pxz}%
  \BibitemOpen
  \bibfield  {author} {\bibinfo {author} {\bibfnamefont {Chiara}\ \bibnamefont {Caprini}}, \bibinfo {author} {\bibfnamefont {Daniel~G.}\ \bibnamefont {Figueroa}}, \bibinfo {author} {\bibfnamefont {Raphael}\ \bibnamefont {Flauger}}, \bibinfo {author} {\bibfnamefont {Germano}\ \bibnamefont {Nardini}}, \bibinfo {author} {\bibfnamefont {Marco}\ \bibnamefont {Peloso}}, \bibinfo {author} {\bibfnamefont {Mauro}\ \bibnamefont {Pieroni}}, \bibinfo {author} {\bibfnamefont {Angelo}\ \bibnamefont {Ricciardone}}, \ and\ \bibinfo {author} {\bibfnamefont {Gianmassimo}\ \bibnamefont {Tasinato}},\ }\bibfield  {title} {\enquote {\bibinfo {title} {{Reconstructing the spectral shape of a stochastic gravitational wave background with LISA}},}\ }\href {\doibase 10.1088/1475-7516/2019/11/017} {\bibfield  {journal} {\bibinfo  {journal} {JCAP}\ }\textbf {\bibinfo {volume} {11}},\ \bibinfo {pages} {017} (\bibinfo {year} {2019})},\ \Eprint {http://arxiv.org/abs/1906.09244} {arXiv:1906.09244 [astro-ph.CO]} \BibitemShut {NoStop}%
\bibitem [{\citenamefont {Speagle}(2020)}]{Speagle:2019ivv}%
  \BibitemOpen
  \bibfield  {author} {\bibinfo {author} {\bibfnamefont {Joshua~S.}\ \bibnamefont {Speagle}},\ }\bibfield  {title} {\enquote {\bibinfo {title} {{dynesty: a dynamic nested sampling package for estimating Bayesian posteriors and evidences}},}\ }\href {\doibase 10.1093/mnras/staa278} {\bibfield  {journal} {\bibinfo  {journal} {Mon. Not. Roy. Astron. Soc.}\ }\textbf {\bibinfo {volume} {493}},\ \bibinfo {pages} {3132--3158} (\bibinfo {year} {2020})},\ \Eprint {http://arxiv.org/abs/1904.02180} {arXiv:1904.02180 [astro-ph.IM]} \BibitemShut {NoStop}%
\bibitem [{\citenamefont {Ashton}\ \emph {et~al.}(2019)\citenamefont {Ashton} \emph {et~al.}}]{Ashton:2018jfp}%
  \BibitemOpen
  \bibfield  {author} {\bibinfo {author} {\bibfnamefont {Gregory}\ \bibnamefont {Ashton}} \emph {et~al.},\ }\bibfield  {title} {\enquote {\bibinfo {title} {{BILBY: A user-friendly Bayesian inference library for gravitational-wave astronomy}},}\ }\href {\doibase 10.3847/1538-4365/ab06fc} {\bibfield  {journal} {\bibinfo  {journal} {Astrophys. J. Suppl.}\ }\textbf {\bibinfo {volume} {241}},\ \bibinfo {pages} {27} (\bibinfo {year} {2019})},\ \Eprint {http://arxiv.org/abs/1811.02042} {arXiv:1811.02042 [astro-ph.IM]} \BibitemShut {NoStop}%
\bibitem [{\citenamefont {Romero-Shaw}\ \emph {et~al.}(2020)\citenamefont {Romero-Shaw} \emph {et~al.}}]{Romero-Shaw:2020owr}%
  \BibitemOpen
  \bibfield  {author} {\bibinfo {author} {\bibfnamefont {I.~M.}\ \bibnamefont {Romero-Shaw}} \emph {et~al.},\ }\bibfield  {title} {\enquote {\bibinfo {title} {{Bayesian inference for compact binary coalescences with bilby: validation and application to the first LIGO\textendash{}Virgo gravitational-wave transient catalogue}},}\ }\href {\doibase 10.1093/mnras/staa2850} {\bibfield  {journal} {\bibinfo  {journal} {Mon. Not. Roy. Astron. Soc.}\ }\textbf {\bibinfo {volume} {499}},\ \bibinfo {pages} {3295--3319} (\bibinfo {year} {2020})},\ \Eprint {http://arxiv.org/abs/2006.00714} {arXiv:2006.00714 [astro-ph.IM]} \BibitemShut {NoStop}%
\bibitem [{\citenamefont {Boileau}\ \emph {et~al.}(2022)\citenamefont {Boileau}, \citenamefont {Jenkins}, \citenamefont {Sakellariadou}, \citenamefont {Meyer},\ and\ \citenamefont {Christensen}}]{Boileau:2021gbr}%
  \BibitemOpen
  \bibfield  {author} {\bibinfo {author} {\bibfnamefont {Guillaume}\ \bibnamefont {Boileau}}, \bibinfo {author} {\bibfnamefont {Alexander~C.}\ \bibnamefont {Jenkins}}, \bibinfo {author} {\bibfnamefont {Mairi}\ \bibnamefont {Sakellariadou}}, \bibinfo {author} {\bibfnamefont {Renate}\ \bibnamefont {Meyer}}, \ and\ \bibinfo {author} {\bibfnamefont {Nelson}\ \bibnamefont {Christensen}},\ }\bibfield  {title} {\enquote {\bibinfo {title} {{Ability of LISA to detect a gravitational-wave background of cosmological origin: The cosmic string case}},}\ }\href {\doibase 10.1103/PhysRevD.105.023510} {\bibfield  {journal} {\bibinfo  {journal} {Phys. Rev. D}\ }\textbf {\bibinfo {volume} {105}},\ \bibinfo {pages} {023510} (\bibinfo {year} {2022})},\ \Eprint {http://arxiv.org/abs/2109.06552} {arXiv:2109.06552 [gr-qc]} \BibitemShut {NoStop}%
\end{thebibliography}%
\end{document}